%% file: ex_article.tex
\documentclass[hidelinks,onefignum,onetabnum]{siamart220329}

\input{ex_shared}

\ifpdf
\hypersetup{
  pdftitle={Nested Sampling and Rare Event Estimation},
  pdfauthor={J. Latz, D. Schneider, and P. Wacker}
}
\fi

\begin{document}

\maketitle

\begin{abstract}
Nested Sampling is a method for computing the Bayesian evidence, also called the marginal likelihood, which is the integral of the likelihood with respect to the prior. More generally, it is a numerical probabilistic quadrature rule. The main idea of Nested Sampling is to replace a high-dimensional likelihood integral over parameter space with an integral over the unit line by employing a push-forward with respect to a suitable transformation. Practically, a set of active samples ascends the level sets of the integrand function, with the measure contraction of the super-level sets being statistically estimated. We justify the validity of this approach for integrands with non-negligible plateaus, and demonstrate Nested Sampling's practical effectiveness in estimating the (log-)probability of rare events.
\end{abstract}

\begin{keywords}
Nested Sampling, Marginal Likelihood, Quadrature, Rare Event Estimation 
\end{keywords}

\begin{MSCcodes}
65C05, 28A25, 62-08
\end{MSCcodes}

\section{Introduction}
Throughout this work, we study the numerical approximation of  expected values
\begin{equation}\label{eq:main}
    \mu(\L) := \int_\Omega \L(x) \diff\mu(x)
\end{equation}

where $(\Omega, \mathcal A, \mu)$ with $\Omega\subset \R^d$ is a probability space and  $\L :\Omega\to \R$ is a measurable function. We are thinking especially of cases where the evaluation of $\L$ is computationally costly, e.g., requiring numerical simulations of partial differential equations or stochastic processes. Here, our methods are particularly useful, if 
\begin{itemize}
    \item[(i)] we are able to sample independently from $\mu$, but a usual Monte Carlo approximation with samples from $\mu$ would be inefficient, e.g., where $\L$ and $\mu$ have most of their masses concentrated in different regions of $\Omega$ or
    \item[(ii)] we need to compute $\mu(\L_\theta)$ for a large number of similar $\L_\theta$, where $\Theta \ni \theta$ is an appropriate parameter space. Similarity is here expressed in terms of the level sets of the functions, assuming that $\{\{\L_\theta = \alpha\} : \alpha \in \mathbb{R}\} \approx \{\{\L_\zeta = \alpha\} : \alpha \in \mathbb{R}\}$, for $\theta, \zeta \in \Theta$, where we use the usual way of denoting preimages $\{\L \in A\} := \{\omega \in \Omega: \L(\omega) \in A\} $ for any $A \subseteq \mathbb{R}$. 
\end{itemize}
We give several examples for quadrature problems of the forms (i) and (ii) in the next section.

\subsection{Overview of this manuscript}

To illustrate the importance of computationally efficient methods for integrals of type \eqref{eq:main}, we now give examples of their application in uncertainty quantification, statistics, physics, and engineering.

\paragraph{Model evidence}
In Bayesian statistics, we encounter integrals of the described form $\mu(\L)$ when computing the \emph{(model) evidence} or \emph{marginal likelihood}. Here, $\L(x) = L(y|x)$, where $L$ describes the likelihood of the data set $y$ given the parameter $x \in \Omega$; $\mu$ is the prior distribution. The model evidence is a vital tool in Bayesian model selection \cite{mackay2003information}, but hard to compute, if the likelihood is highly concentrated. A closely related problem in statistical physics is that of estimating the partition function of a system in the thermodynamic equilibrium.

\paragraph{Rare event estimation} We consider an event $G \in \mathcal{A}$ with $\mu(G)$ small, say, $\approx 10^{-9}$. Of course, $\mu(G)=\mu(\L),$ in the setting above, when choosing $\L$ to be the indicator or characteristic function $\chi_G$; defined by  $\chi_G(\omega) = 1,$ if $\omega \in G$ and $\chi_G(\omega) = 0,$ otherwise. Rare event estimation is a central tool in reliability analysis, where failures of physical system need to be estimated, but also in finance. A straightforward Monte Carlo approximation of a rare event requires a number of samples of order $p^{-1}$ -- unachievable in most applications; especially when physical simulations are necessary.

\paragraph{Moments and the moment-generating function} The moment generating function of $\L$ is defined as $M_H(\theta) = \int_{\Omega}\exp(\theta H(x))\mathrm{d}\mu(x)$, where $H: \Omega \rightarrow \mathbb{R}$ is a random variable and $\theta \in \Theta = \mathbb{R}$. The function $M_H$ can be used to accurately represent the probability distribution of the random variable $H(X)$, where $X \sim \mu$. Here, $\L_\theta := \exp(\theta H)$ satisfies the condition mentioned above under (ii). Either through the moment generating function or using the immediate condition, we are also able to compute all moments of $H(X)$. Here, we can choose $\L_\theta = H^\theta$, for $\theta \in \mathbb{N} := \{1,2,\ldots\}$.

\paragraph{Cumulative distribution functions and quantiles} The estimation of the cumulative distribution function (cdf) $F := \mu(Y \leq \cdot)$ of a random variable $Y: \Omega \rightarrow \mathbb{R}$ is closely related to the rare event estimation problem, with the difference that we are now interested in the probability of the event $\{Y \leq y\}$ for a range of values $y \in \mathbb{R}$ (with $y\to - \infty$ corresponding to increasingly rare events). A particularly interesting problem is the estimation of quantiles. Here, we need to find  $q \in \mathbb{R}$ where $F(q) = \alpha$ for a pre-determined $\alpha > 0$. To find $q$, a bisection method may require several evaluations of  $\mu(Y \leq y)$ with similar sets of the form $\{Y \leq y\}$.

\bigskip
For all these integration tasks, we propose application of the celebrated \emph{Nested Sampling} methodology.
Nested Sampling was introduced by \cite{skilling2006nested} as a specialized algorithm for the computation of the Bayesian model evidence  and has been applied successfully in astronomy and computational physics (\cite{baldock2017constant, feroz2008multimodal, feroz2009multinest, mukherjee2006anested,murray2006nested,parkinson2013bayesian,partay2014nested,vegetti2009bayesian,veitch2015parameter}), biomathematics (\cite{aitken2013nested, dybowski2013nested, pullen2014bayesian}), and other fields. 
From a bird's eye view, Nested Sampling is a method of computing the integral $\int_\Omega \L \diff \mu$ by converting this high-dimensional integral into an integral on the unit line $[0,1]$:\footnote{The point of view of \cite{skilling2006nested} is slightly different, interpreting $\int_\Omega \mathcal{L}(x) \diff \mu(x) = \int_0^1 \tilde{\mathcal{L}}(X) \diff X$, where $\tilde\L$ is an ``overloaded'' form of the likelihood $\L$ but we will work with \eqref{eq:Nested-sampling-ours}, since it allows us to generalise the methodology considerably. 
}
\begin{equation} \label{eq:Nested-sampling-ours}
     \int_\Omega \L(x) \diff \mu(x) \stackrel{(\clubsuit)}= -\int_0^1 \lambda \diff X(\lambda) \stackrel{(\heartsuit)}\approx  \sum_{i=1}^N \lambda_i \cdot \left(X_{i-1}-X_{i} \right) \stackrel{(\spadesuit)}\approx\sum_{i=1}^N\lambda_i\cdot D_i,  \end{equation}
     where $X(\lambda)=\mu(\L > \lambda)$ is the \textit{survival function} of $X$ and the $(X_i)_{i=1}^N$ are a suitable discretization $X_i = X(\lambda_i)$ derived further below.
     This means that the integral in question is
     \begin{enumerate}
         \item[$(\clubsuit)$] converted into a Riemann-Stieltjes-integral first, 
         \item[$(\heartsuit)$] which is approximated by a specific (randomized) quadrature rule,
         \item[$(\spadesuit)$] with the probabilistic \textit{contraction} ($X_{i-1}-X_i$) being replaced by a deterministic estimator $D_i$.
     \end{enumerate}
        In some sense, Nested Sampling can be thought of as a Lebesgue-integral version of stochastic quadrature, in the same way that usual Monte Carlo quadrature corresponds to Riemann integration (with probabilistically chosen quadrature points). This point of view becomes clearer from the proof of \Cref{lem:club} below.
Here, we use a set of particles to approximate increasing (nested) super-level sets of $\L$. The contraction of the measure of those super-level sets is then approximated by a deterministic shrinkage estimator.
    
There is already an extensive body of work regarding the performance of the second step $(\heartsuit)$, i.e.\ the quality of the Monte Carlo approximation of the one-dimensional integral 
and a series of improvements have been proposed \cite{higson2018sampling,feroz2013importance,higson2019dynamic,salomone2018unbiased}. To the best of our knowledge there is a gap in the justification of steps $(\clubsuit)$ and $(\spadesuit)$, i.e.\ the substitution of the integral over $\Omega$ with an integral over the unit interval, and the validity of the deterministic approximation. This manuscript tries to close this gap in the literature. The crucial issue allowing Nested Sampling to be applied for general integration tasks (especially those outlined in (i) and (ii)), is the integration of functions with plateaus. We say, $\mathcal{L}$ has a plateau, if there is some $c \in \mathbb{R}$, for which the level set $\{\mathcal{L} = c\}$ has positive Lebesgue measure.

Computational problems with using, in particular, integrands with non-vanishing plateaus have been described by \cite{murray2007advances} and \cite{skilling2006nested}, who suggest a randomization or ``labelling'' approach, breaking a tie between points with the same likelihood value. We argue in this manuscript that the version proposed by \cite{fowlie2020nested} is a computationally more suitable way for handling possible plateaus in the likelihood. While this manuscript did not provide a proof of correctness, and it is indeed not trivial to do so, we give some justification for the correctness of this modification of Nested Sampling. 

The contributions of this manuscript are, stated succinctly, as follows:

\begin{enumerate}[leftmargin=2.5cm]
    \item[\Cref{sec:NewAttemptAtExplanationOfNestedSampling}] describes the idea of Nested Sampling and shows that the integral transformation $(\clubsuit)$ given by $\int_\Omega \mathcal{L}(x) \diff \mu(x) = -\int_0^1 \lambda \diff X(\lambda)$ is valid even for very general functions $\mathcal{L}$ with plateaus. This is achieved by viewing Nested Sampling as a numerical quadrature of a Riemann-Stieltjes-integral, a perspective that apparently has not been explored so far. 
    \item[\Cref{sec:nestedsampling}] justifies with some examples why the modification proposed by \cite{fowlie2020nested} is a good idea in order to make sure that the deterministic approximation $(\spadesuit)$ is valid. This is not entirely straightforward to see and depends on slightly obscure properties of survival functions. Essentially, the deterministic estimate for the contraction $X(\lambda_i)-X(\lambda_{i+1})$ of the available prior mass has to be justified.
    \item[\Cref{sec:UQ_REE}] proposes a way of unifying large classes of integrals of the form \eqref{eq:main} by means of \textit{level set surrogates}, allowing efficient application of Nested Sampling to rare event estimation, approximation of the survival function of a random variable, and simultaneous estimation of higher-order moments.
    \item[\Cref{sec:simulations}] demonstrates the performance of Nested Sampling in the tasks outlined in \Cref{sec:UQ_REE}.
\end{enumerate}

\subsection{The paradigm of Nested Sampling}\label{sec:NewAttemptAtExplanationOfNestedSampling}
We begin with a quick stand-alone explanation of Nested Sampling, i.e. how exactly \eqref{eq:Nested-sampling-ours} can be derived. This derivation is partially restricted (for purely pedagogical reasons) to the case that $\L$ does not have any plateaus of non-zero measure. We will later drop this assumption.
\begin{assumption}\label{ass:mainNeSa} We consider a probability space $(\Omega, \mathcal A, \mu)$. Let $\L:\Omega\to \R$ be a measurable function. We define $\mu^{\lambda}$ as the conditional measure given  $\{\L > \lambda\}$, i.e. $ \mu^\lambda(A):= \frac{\mu(A \cap \{\L > \lambda\})}{\mu(\L > \lambda)}$  and we assume that
    \begin{myenum}
        \item \textbf{Positivity:} $\L > 0$ almost surely, and
        \item \textbf{Finite Moments:} $\E \L^p < \infty$ for some $p>1$.
    \end{myenum}
\end{assumption}

    Note that this also works with any $\hat\L$ bounded from below by any $C\in\R$, which can be reduced to the case $\L > 0$ by setting $\L = \hat \L - C$ and seeing that $\int \hat \L\diff\mu = \int \L\diff\mu + C$. 

\begin{lemma}[validity of $\clubsuit$]\label{lem:club}
    If \Cref{ass:mainNeSa} holds, we have
    \[\int_\Omega \L(x) \diff\mu(x)\stackrel{\clubsuit}=  \int_0^\infty \lambda \diff (-X(\lambda)),\]
    where $X(\lambda) := \mu(\L > \lambda)$ is the so-called \emph{survival function} of $\L$.
\end{lemma}
\begin{proof}
We define $X(\lambda)$ as stated above and see that
\begin{align*}
    \int_\Omega \L(x) \diff\mu(x) &= \E^\mu \L = \int_0^\infty \mu(\L > \lambda) \diff \lambda = \int_0^\infty X(\lambda)\diff \lambda.
    \intertext{The Riemann-Stieltjes integral $\int_0^r \lambda \mathrm d X(\lambda)$ is well-defined and we can rewrite}
   \int_0^\infty X(\lambda)\diff \lambda &= \lim_{r\to\infty}\int_0^r X(\lambda)\diff \lambda = \lim_{r\to\infty} X(r)r - X(0)0 + \lim_{r\to\infty} \int_0^r \lambda \diff (-X(\lambda)).
    \intertext{By Markov's inequality,  $0 \leq \lim_{r\to\infty}X(r)r \leq \lim_{r\to\infty}\E\L^p r^{1-p} = 0$. Then} 
 \lim_{r\to\infty} &X(r)r - X(0)0 + \lim_{r\to\infty} \int_0^r \lambda \diff (-X(\lambda))   =  \int_0^\infty \lambda \diff (-X(\lambda)).\qquad
\end{align*}
\end{proof}

\paragraph{Approximating the one-dimensional integral ($\heartsuit$)}
By definition, Riemann-Stiel\-tjes integrals can be approximated via
\begin{align*}
 \int_0^\infty \lambda \diff (-X(\lambda))\approx \sum_{i=0}^{N-1} \lambda_{i+1} \left[X(\lambda_{i}) - X(\lambda_{i+1}) \right]
\end{align*}
for some choice of $0 = \lambda_0 < \lambda_1 < \cdots < \lambda_N$ where we set $X(\lambda_0) = 1$. Unfortunately, if we just set some fixed points $\lambda_i$, we do not have any straight-forward way of calculating $\left[X(\lambda_{i-1}) - X(\lambda_i) \right] = \mu(\L\in (\lambda_{i-1}, \lambda_i])$. The trick of Nested Sampling is now to find a clever way of choosing the points $\lambda_i$ such that the quantity $X(\lambda_{i-1}) - X(\lambda_i)$ can be statistically estimated. This is the content of \Cref{lem:contraction_noplateau}, which relies on the following additional assumption to hold.

\begin{assumption}\label{ass:mainNeSaStronger} We assume that \Cref{ass:mainNeSa} holds, and additionally
    \begin{myenum}
        \item \textbf{No plateaus:} $\mu(\L = \lambda) = 0$ for all $\lambda \in \R$ (equivalently, the cumulative distribution function of $\L$ is continuous), and \label{ass:noplateau}
        \item \textbf{Super-level sampling:} we can efficiently sample from $\mu^\lambda$. \label{ass:sampling}
    \end{myenum}
\end{assumption}
    \Cref{ass:mainNeSaStronger}\Cref{ass:sampling} seems (and often is) restrictive, but is in practice fulfilled in the context of Nested Sampling, as we will see further below.

We start with $\lambda_0 = \inf \L = 0$, i.e. $\mu^{\lambda_
0} = \mu$. We now describe how to inductively choose $\lambda_{i+1}$ from given $\lambda_i$: We generate independent and identically distributed (i.i.d.) samples $\{x^k\}_{k=1}^J\sim \mu^{\lambda_i}$, which is possible according to \Cref{ass:mainNeSa}\Cref{ass:sampling}. We then set 
\begin{equation}
    \label{eq:newlambda} \lambda_{i+1} = \min_{k=1}^J \L(x^k)
\end{equation}
and
\begin{equation}
    \xi_{i+1} := \max_k X(\L(x^k)) = X(\lambda_{i+1}).
\end{equation}
Equality holds due to the fact that  $\argmin_k \L(x^k) = \argmax_k X(\L(x^k))$ because $X$ is non-increasing. The next lemma characterizes what we can say about the \textit{contraction} $X(\lambda_i) - X(\lambda_{i+1})$.
\begin{lemma}[validity of $\spadesuit$, no-plateau case]\label{lem:contraction_noplateau}
    If \Cref{ass:mainNeSaStronger} holds, then the following statements are true.
    \begin{enumerate}
        \item If $\{x^k\}_k\sim \mu$, then $\{X(\L(x^k))\}_k\sim U[0,1]$. The random variable $\xi_1 := \max_k X(\L(x^k))$ follows a $\mathrm{Beta}(J,1)$ distribution with mean 
        \[\E^{x^k\sim \mu} \left[ \max_k X(\L(x^k))\right] = \frac{J}{J+1}\]
        Writing $\lambda_1 := \min_k\L(x^k)$, i.e., $X(\lambda_1) = \xi_1$, we have
        \[\E^{x^k\sim \mu} \left[ 1 - \xi_1\right] = \frac{J}{J+1}.\]
        \item Recursively in $i\geq 1$: Consider $\lambda_i \geq 0$ and $\xi_i$. 
        If $\{x^k\}_k\sim \mu^{\lambda_i}$,\footnote{i.e. $x^k\sim \mu$ under the additional condition $\L(x^k) > \lambda_i$} then $\{X(\L(x^k))\}_k \sim U[0,\xi_i]$. The random variable $\xi_{i+1} := \max_k X(\L(x^k))$ follows a rescaled $\mathrm{Beta}(J,1)$ distribution on the interval $[0,\xi_i]$ with mean 
    \[\E^{x^k\sim \mu^{\lambda_i}} \left[\max_k X(\L(x^k))\right] = \frac{J}{J+1}\xi_i.\] Writing $\lambda_{i+1} := \min_k\L(x^k)$, i.e., $X(\lambda_{i+1}) = \xi_{i+1}$,\footnote{Note that only the term $\xi_{i+1}$ is averaged over inside the square brackets, $\xi_i$ being fixed from the last iteration.}
    \[\E^{x^k\sim \mu^{\lambda_i}}\left[ \xi_i - \xi_{i+1} \right]= \frac{1}{J+1}\xi_i.\]
    \item Defining $\lambda_0 := 0$ and $\xi_0 = 1$ for consistency,
     \[\E \left[\xi_i-\xi_{i+1} \right] = \frac{J^i}{(J+1)^{i+1}},\]
     where the expectation are over recursive repeated sampling of $\{x^k\}\sim\mu^{\lambda_i}$ for $i=0,\ldots,i-1$.
    \end{enumerate}
\end{lemma}

    The last statement of this lemma motivates the form of the deterministic estimator $D_i = \frac{J^i}{(J+1)^{i+1}}$ in \eqref{eq:Nested-sampling-ours}$(\spadesuit)$.
\begin{proof}
We define $\F_\L$ as the cumulative distribution function of $\L$. According to \cite[proposition 2.(1)]{embrechts2013note}, or \Cref{lem:uniformmeasure}, $(\F_\L\circ \L)_\#\mu = U[0,1]$.\footnote{This means that if we sample $\{x^k\}_k$ from $\mu$, then the transformed samples $\{F_\L(\L(x^k))\}_k$ will be uniformly distributed on $[0,1]$.} We now compute\footnote{I.e. if we sample $\{x^k\}_k$ from $\mu^\lambda$, what is the distribution of $\{F_\L(\L(x^k))\}_k$?} $(\F_\L\circ \L)_\#\mu^\lambda$ by analysing its cumulative distribution function. Note that $\mu^\lambda(A) = \mu(A \cap \{\L > \lambda\})/\mu(\L > \lambda)$. For brevity of exposition, we interpret $\F_\L\circ \L$ directly as a random variable, this is to be understood in the sense that we consider the measure $(\F_\L\circ \L)_\#\mu$. 

Now, using the fact that the inverse of the cumulative distribution function $\F_\L$ is well-defined due to \Cref{ass:mainNeSaStronger}\cref{ass:noplateau},
\begin{align*}
\mu^\lambda&(\F_\L\circ \L \leq r) = \frac{\mu(\F_\L \circ \L \leq r \text{ and } \L > \lambda)}{\mu( \L > \lambda)}\\
&= \frac{\mu(\F_\L \circ \L \leq r \text{ and } \F_\L\circ \L > \F_\L(\lambda))}{1-\F_\L(\lambda)}\\
&= \frac{\mu(\F_\L \circ \L \leq r) - \mu(\F_\L \circ \L \leq \F_\L(\lambda))}{1-\F_\L(\lambda)}\\
&= \frac{r-\F_\L(\lambda)}{1-\F_\L(\lambda)} \text{ for $r\in [\F_\L(\lambda),1]$ and $0$ otherwise,}
\end{align*}
where the last step is due to the fact that $(\F_\L\circ \L)_\#\mu = U[0,1]$, i.e. $ \mu(\{x\in\Omega: \F_\L(\L(x))\leq r\}) = r$ for $r\in[0,1]$. This means that $(\F_\L\circ \L)_\#\mu^\lambda = U[\F_\L(\lambda),1]$.

Now we use the fact that $X(\lambda) = 1 - \F_\L(\lambda)$, and thus $(X\circ \L)_\#\mu^\lambda = U[0, 1 - \F_\L(\lambda)] = U[0,X(\lambda)]$. The remaining statements follow directly from the fact that the maximum of $J$ uniform distributions is a Beta distribution, recursive reasoning, and elementary computation.
\end{proof}

After this derivation of the variable substitution at the heart of Nested Sampling (in the restricted case of \Cref{ass:mainNeSaStronger}) we want to point out a few computational techniques relevant to Nested Sampling.

\paragraph{Concrete implementation and computational tricks}
Nested Sampling performs a kind of noisy Riemann-Stieltjes integration: We compute
\[\int_\Omega \L(x) \diff\mu(x)=  \int_0^\infty \lambda \diff (-X(\lambda))\approx \sum_{i=0}^{N-1} \lambda_{i+1} \left[X(\lambda_{i}) - X(\lambda_{i+1}) \right] \]
where we can only infer $\left[X(\lambda_{i}) - X(\lambda_{i+1}) \right]$ statistically, e.g. by approximating it by the unbiased estimator $\E \left[ X(\lambda_i) - X(\lambda_{i+1}) \right] = \frac{J^i}{(J+1)^{i+1}}$ obtained from \Cref{lem:contraction_noplateau}.

A relevant computational ingredient is the use of the \texttt{logsumexp} function in order to avoid numerical underflow: The values of $\lambda_i$ are increasing in $i$, while $X(\lambda_{i}) - X(\lambda_{i+1})$ is set to a fixed exponentially decreasing sequence of the form $\frac{J^i}{(J+1)^{i+1}}$. This means that for most terms in this sum, either of those terms will be very small. In addition, the value of $\int \L\diff\mu$ is in most applications almost astronomically small, e.g. when $\L$ is a likelihood in a high-dimensional Bayesian inverse problem with relatively highly informative data, or when $\L$ is the characteristic function of an event with very small probability. For this reason, we usually compute $\log\int \L\diff\mu$ instead of $\int \L\diff\mu$. The \texttt{logsumexp} function provides a way of computing $\log(\sum_i \exp(a_i))$ iteratively without loss of information. That this is indeed relevant can be seen with the following simple example: Set $a_1 = -800$ and $a_2 = -801$. Then naive implementation $\log(\exp(a_1) + \exp(a_2))$ would lead to an error since both terms in the sum would be evaluated to $0$ in most floating point arithmetics, and inserted into the logarithm. On the other hand, we can write $\log(\exp(a_1) + \exp(a_2)) = \log(\exp(a_1)(1 + \exp(a_2-a_1))) = a_1 + \log(1+\exp(a_2-a_1))$ which can be evaluated without any problem. This way we can iteratively compute the logarithmic sum $\log \sum_{i=0}^{N-1} \lambda_{i+1} \left[X(\lambda_{i}) - X(\lambda_{i+1}) \right]$ without incurring too much loss of accuracy.

A seemingly major stumbling block for the implementation of Nested Sampling is the issue of \Cref{ass:mainNeSa}\Cref{ass:sampling}: How do we sample from $\mu$ under the additional assumption that $\L > \lambda_{i}$? This is not too big a challenge in practice and we now describe how this can be done in a recursive manner over the iterations $i$. The case $i=1$ is trivial, since $\mu^{\lambda_0} = \mu$, i.e. we are sampling from the unconstrained measure $\{x^k\}_{k=1}^J\sim \mu$. We pick $\lambda_1 = \min_k \L(x^k)$, and estimate $X(\lambda_0) - X(\lambda_1) = \frac{1}{J+1}$. This gives us our first term in the sum, $\lambda_1 \cdot (X(\lambda_0) - X(\lambda_1) )$ and we can increment the loop counter $i$. Now we would have to sample from $\mu^{\lambda_1}$, which is a priori a nontrivial task. But the samples from the previous iteration carry information that we can use: Since $\lambda_1 = \min_k \L(x^k)  < \L(x^l)$ for $l\neq \argmin_k \L(x^k)$, all but one of the old samples give us a good idea of where to look since they already are samples from $\mu^{\lambda_0}$ fulfilling $\L > \lambda_1$, and therefore, are samples from $\mu^{\lambda_1}$. We just need to remove one sample (the particle $x^k$ with $k = \argmin_k \L(x^k)$) and replace it by a sample from $\mu^{\lambda_1}$. But since we already have $J-1$ samples from this measure, we can use an ``explorative'' sampler like Slice Sampling or a Metropolis--Hastings variant to start there and obtain a new sample from $\mu^{\lambda_1}$. In fact, we found that a good explorative method is critical for Nested Sampling's performance in most applications we considered.

\section{Nested Sampling: The plateau case}\label{sec:nestedsampling}

Nested Sampling was originally introduced as a methodology to compute model evidences in Bayesian inference. In this case, the function $\L$ is a likelihood. Likelihood functions $\L$ with plateaus, i.e. $\L$ having level sets with positive prior measure, appear only infrequently in the literature. There, we find them when considering rounded data or general piecewise constant statistical models\footnote{In the context of quantization in electrical engineering, signals are rounded to a grid. The likelihood of the true signal given such a quantized version is then a uniform distribution over the range of numbers being rounded to this value \cite{sripad1977necessary,bjorsell2007truncated}. They also appear in more general settings where, e.g., piecewise constant parameter-data dependencies occur, see, e.g., \cite{Latz2020}.} that lead to a flat plateau at the likelihood's peak. In other scenarios, the likelihood is supported on a compact domain, e.g. $\{\L = 0\}$ has positive measure\footnote{Some applications come with explicit knowledge about upper bounds on the measurement error's magnitude which corresponds to a plateau (of magnitude 0) on the range of impossible original signals (e.g. in biological applications; in the context of image processing on grayscale images with values in the unit interval; or when reading out an analog thermometer by looking at the nearest labeled tick on the scale: Here, the maximum measurement error is the distance between adjacent ticks on the scale). In \cite{Yao2017} the authors argue that bounded noise is more realistic in some biological and physical context and that the choice of the correct noise model has a large influence on the long-term behaviour of models. In econometrics (e.g. \cite{glewwe2007measurement}), lognormal measurement noise is sometimes used as a tractable way of modelling noise terms guaranteeing non-negative data. See also \cite{d2013bounded} and the references therein for more examples of bounded noise in physics, biology, and engineering.}.

\begin{figure}
    \centering
    \includegraphics[width=0.8\textwidth]{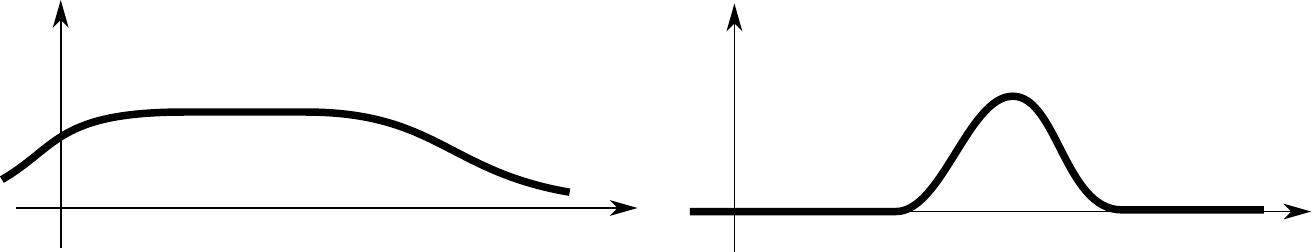}
    \caption{Left: Plateau as a peak. Right: Plateau by compact support.}
    \label{fig:plateaus}
\end{figure}

However, we explicitly consider the plateau case in this work as we are interested in much more general integration tasks that go beyond model evidences. When estimating the probabilities of rare events, $\L$ is an indicator function and, thus, comprised of exactly two plateaus. Integrands with plateaus also appear in the estimation of risk measures, such as the conditional value at risk that has recently gained significance in stochastic control problems \cite{Miller}. The statistical-physics-equivalent of estimating the model evidence in Bayesian inference is the computation of the partition function. Here, piecewise-constant integrands arise through piecewise-constant potentials in, e.g., molecular dynamics \cite{Kasperskovitz05} and statistical thermodynamics \cite{walker}. Outside of these specific, computationally expensive problems, the general integration of piecewise-constants functions may already be of interest. If the positions of the discontinuities is not known, many classical quadrature rules may fail as they usually rely on smoothness in the integrand and its precise approximation through polynomials \cite{Trefethen}.

We showed the validity of \eqref{eq:Nested-sampling-ours}, i.e. of steps $(\clubsuit)$, $(\heartsuit)$, $(\spadesuit)$ in the no-plateau case. The first equation ($\clubsuit$) was proven in a way which does not depend on whether the integrand has a plateau or not. Equation ($\heartsuit$) is just the discretization of the Riemann--Stieltjes integral, so it remains to investigate whether the deterministic estimator in ($\spadesuit$) is a reasonable approximation even in the plateau case.

\subsection{Correctness of the contraction estimate for integrands with non-vanishing plateaus}

The first proof of correctness of Nested Sampling was given in \cite{evans2007discussion}, but does not consider the problematic case of discontinuous survival functions (corresponding to plateaus in the integrand, see below). Similarly, \cite{salomone2018unbiased} gives a very short and precise derivation via the inverse of the survival function, but again only for integrands without plateaus of non-negligible measure. In \cite{chopin2010properties}, the authors refer to \cite{burrows1980new} for justification of the integral transformation, but in our opinion the issue is quite difficult and deserves a more careful analysis. At the heart of the matter, Nested Sampling amounts to integration with respect to the push-forward measure, and the basic idea hinges on properties of survival functions and generalized inverses which are not readily found explicitly in the existing literature.

The main difficulty arises when $\L$ has a non-negligible plateau
were already hinted at (although in the context of the algorithm's performance) in the original publication \cite{skilling2006nested} with a discussion of ``cliffs'' -- deemed non-problematic -- and ``plateaus''. The author recognizes the difficulty of plateaus (for reasons laid out below), but rules ``[...] even so, it may be possible to generate [new active samples] efficiently.'' The PhD thesis \cite{murray2007advances} also mentions this issue and further improves on Skilling's original suggestion.

We will show that plateaus in ${\L}$ are not only computationally troublesome but also fundamentally mathematically problematic: If there is a level $\alpha \in \R$ such that $\mu(\L = \alpha) > 0$, then the following implicit uniformity assumption about Nested Sampling is violated:
``Samples from the prior which are plotted in a $X$-$\L$-diagram are uniformly distributed along the axis $[0,1]$.''  This is what the authors of \cite[section 9.2]{sivia2006data} mean when they write ``In terms of $\xi$, the objects are uniformly sampled subject to the constraint $\xi < \xi^\star$''. This means that the approximation $(\spadesuit)$ in \eqref{eq:Nested-sampling-ours} is invalid, because the uniformity assumption of $X\circ \L$ in \Cref{lem:contraction_noplateau} no longer holds.

This uniformity assumption (which is conditional to $\L$ not having plateaus of non-vanishing mass) is crucial for the statistical contraction estimate empowering Nested Sampling, and is assumed to hold in every exposition of Nested Sampling known to us, as with the original manuscript \cite[section 5]{skilling2006nested}, but also in expositions since then, \cite[section 3]{evans2007discussion}, \cite[section 2.1]{keeton2011statistical}, \cite[section 2]{higson2018sampling}, to just name a few, usually derived from the statistics of the largest of $M$ uniform deviates, where $M\in\N$ is the number of active samples.

In the context of the applications considered in this manuscript, the integrand $\L$ will usually have a plateau, and thus we need a theory of what to do in this case. Fortunately, we can prove that this mathematical problem can be circumvented. Lemma \ref{lem:formofXL} shows that $\F_\L\circ \L$ (and thus, $X\circ \L$) is indeed \textit{not} uniformly distributed, but gives the correct form of this measure, which is, essentially, a well-behaved combination of uniform distributions and Dirac masses.

We recall the following well-known properties of cumulative distribution functions:
\begin{lemma}
We consider a probability space $(\Omega, \mathcal A, \mu)$ and a measurable function $\L:\Omega\to \R$. Then the cumulative distribution function of $\L$ given by $\F_\L(\lambda) = \mu(\L\leq \lambda)$ is a non-decreasing c\`adl\`ag function, i.e. 
\begin{itemize}
    \item \textbf{Nondecreasing property:} For any $\lambda_1 < \lambda_2$, we have $\F_\L(\lambda_1) \leq \F_\L(\lambda_2)$.
    \item \textbf{Continuity from the right}: For all $\lambda$, $\lim_{\eps\searrow 0} \F_\L(\lambda+\eps) = \F_\L(\lambda)$.
    \item \textbf{Limits from the left}: For all $\lambda$, the limit $\F_\L(\lambda-) := \lim_{\eps\searrow 0} \F_\L(\lambda-\eps)$ is well-defined.
\end{itemize}    
\end{lemma}
Next we characterise plateaus and cliffs of a random variable: A plateau is a level set $\lambda^\star\in \R$ such that $\mu(\L = \lambda^\star) > 0$. This can be characterized as a point in the domain of $\F_\L$ with a lack of left continuity: $\mu(\L = \lambda^\star) = \F_\L(\lambda^\star) - \mu(\F_\L < \lambda^\star)$. Indeed, for $\eps > 0$,
\begin{align*}
    \mu(\F_\L < \lambda^\star) \in \left[\mu(\F_\L \leq \lambda^\star - \eps), \mu(\F_\L \leq \lambda^\star)\right]
\end{align*}
and thus by the sandwich lemma (and letting $\eps\to 0$),
\begin{align*}
    \mu(\F_\L < \lambda^\star) \in \left[\F_\L(\lambda^\star-) , \F_\L(\lambda^\star) \right]
\end{align*}
and
\begin{align*}
    \mu(\L = \lambda^\star) \in [0,  \F_\L(\lambda^\star)-\F_\L(\lambda^\star-) ]
\end{align*}
This means that $\mu(\L = \lambda^\star) > 0$ implies $\F_\L(\lambda^\star) > \F_\L(\lambda^\star-)$. 

A cliff of $\L$ is a pair of levels $\lambda_1<\lambda_2$, such that $\mu(\L\in (\lambda_1,\lambda_2)) = 0$, or equivalently, $\mu(\L\leq \lambda_1) = \mu(\L< \lambda_2)$, i.e. a connected interval $[\lambda_1,\lambda_2)$ such that $\F_\L$ is constant on it. This motivates the following definition.
\begin{definition}
  We consider a probability space $(\Omega, \mathcal A, \mu)$, a measurable function $\L:\Omega\to \R$ and its cdf $\F_\L$. 
   \begin{myenum}
   \item A level $\lambda^\star\in \R$ such that $\F_\L(\lambda^\star) > \F_\L(\lambda^\star-)$ is called a \textbf{plateau} of $\L$. We call $\Delta(\lambda^\star) = \F_\L(\lambda^\star) - \F_\L(\lambda^\star-)$ the \textbf{mass} of the plateau and $\alpha(\lambda^\star) = \F_\L(\lambda^\star-)$ the \textbf{submass} of the plateau.
   \item A \textbf{cliff} of $\L$ is a largest connected interval $I$ of form $[\lambda_1,\lambda_2)$ such that $\F_\L$ is constant on it.\footnote{Note that this does not exclude that $F_\L(\lambda_1)=\F_\L(\lambda_2)$, i.e. that the plateau extends to the full interval $[\lambda_1,\lambda_2]$, this just makes sure that there is no larger interval open to the right of form $[\lambda_1,\lambda_2+\eps)$ on which $\F_\L$ is constant.}
   \end{myenum}
\end{definition}
     Note that if $\lambda^\star$ is not a plateau, then $\Delta(\lambda^\star) = 0$ and $\alpha(\lambda^\star) = \F_\L(\L(\lambda^\star-))$. This further implies that $\alpha(\cdot)$ is the left-continuous version of $\F_\L$.
This now allows us to characterize the exact form of $\F_\L\circ \L$, interpreted as a random variable.
\begin{lemma}\label{lem:formofXL}
We consider a probability space $(\Omega, \mathcal A, \mu)$ and a measurable function $\L:\Omega\to \R$ with the following properties:
   \begin{myenum}
        \item \textbf{Positivity:} $\L > 0$ almost surely,
        \item \textbf{Plateaus:} There is a countable number of plateaus $l_1 < l_2,\ldots$ with non-vanishing mass $\mu(\L = l_i) = \Delta_i$, submass $\mu(\L<l_i) = \alpha_i$, and supermass $\mu(\L>l_i) = \beta_i$. For all $\lambda\not\in \{l_1,\ldots,l_M\}$, we assume $\mu(\L = \lambda) = 0$.\label{ass:plateau}
    \end{myenum}
   Now,  $(\F_\L\circ \L)_\#\mu$ and $(\F_\L\circ \L)_\#\mu^\lambda$ are probability measures on $[0,1]$ characterized by their cumulative distribution in the form
    \begin{align}
        &(\F_\L\circ \L)_\#\mu ([0,\alpha]) = \begin{cases}
            \alpha_i  &\text{ if } \alpha \in[\alpha_i, \alpha_i+\Delta_i)\\
            \alpha  & \text{else.}
        \end{cases}\\
        &(\F_\L\circ \L)_\#\mu^\lambda ([0,\alpha])= \begin{cases}
            \frac{\alpha_i - \F_\L(\lambda)}{1-\F_\L(\lambda)}  &\text{ if } \alpha \in[\alpha_i, \alpha_i+\Delta_i), \alpha > \F_\L(\lambda)\\
            \frac{\alpha - \F_\L(\lambda)}{1-\F_\L(\lambda)}  & \text{ if } \alpha \not\in\bigcup[\alpha_i, \alpha_i+\Delta_i), \alpha > \F_\L(\lambda)\\
            0 & \text{ if }\alpha \leq  \F_\L(\lambda)
        \end{cases}
    \end{align}
    
    If we define $X(\lambda):= 1-\F_\L(\lambda)$,
    \begin{align}
       &(X\circ \L)_\#\mu ([0,\alpha]) =\begin{cases}
            \beta_i + \Delta_i  &\text{ if } \alpha \in[\beta_i, \beta_i+\Delta_i)\\
            \alpha  & \text{else}
        \end{cases}\\
        &(X\circ \L)_\#\mu^\lambda ([0,\beta]) \notag\\&=\begin{cases}
            \frac{\beta_i + \Delta_i}{X(\lambda)}  &\text{ if } \beta \in[\beta_i, \beta_i+\Delta_i) \text{ and } \beta < X(\lambda)\\
            \frac{\beta }{X(\lambda)}  & \text{ if } \beta \not\in\bigcup[\beta_i, \beta_i+\Delta_i) \text{ and } \beta < X(\lambda)\\
            1 & \text{ if }\beta \geq  X(\lambda)
        \end{cases}
    \end{align}
    This means that 
    \begin{align*}
        (\F_\L\circ \L)_\#\mu &= \mathrm{Unif}\left(\bigcup_i [\alpha_i,\alpha_i+\Delta_i)\right)^c + \sum_i \Delta_i \cdot \delta_{\alpha_i+\Delta_i}\\        
        (X\circ \L)_\#\mu &= \mathrm{Unif}\left(\bigcup_i [\beta_i,\beta_i+\Delta_i)\right)^c + \sum_i \Delta_i \cdot \delta_{\beta_i}
    \end{align*}
    where $r\cdot \delta_x$ denotes a Dirac measure of strength $r$ at position $x$, i.e. $\int f(y) \diff(r\cdot \delta_x(y)) = r \cdot f(x)$.
\end{lemma}
\begin{proof}
    In addition to $\F_\L(\lambda) = \mu(\{x \in \Omega: \L(x)\leq \lambda\})$ being the (right-continuous) cumulative distribution function of $\L$, we define its (left-continuous) version $\F_\L^\circ(\lambda) = \mu(\{x \in \Omega: \L(x)< \lambda\})$
    \begin{align*}
        (\F_\L\circ \L)_\#\mu ([0,\alpha)) &= \mu(\{x\in \Omega: \F_\L(\L(x))< \alpha\}) 
        \intertext{Since $\F_\L$ is right-continuous, by \Cref{lem:generalizedinv_prop}\Cref{equiv_rightcont}, we have equal to }
      \mu(\{x\in \Omega: \F_\L(\L(x))< \alpha\})   &= \mu(\{x\in \Omega: \L(x)< \F_\L^-(\alpha)\}) \\
         &= \F_\L^\circ(\F_\L^-(\alpha)) \\
         &= \F_\L^\circ((\F_\L^\circ)^-(\alpha)) \\
        &=\begin{cases}
            \alpha_i  &\text{ if } \alpha \in (\alpha_i, \alpha_i+\Delta_i]\\
            \alpha & \text{else}
        \end{cases}
    \end{align*}
    by an application of \Cref{lem:TTpmTpmT}, and using the fact that the generalized inverse does not depend on the continuity of the function considered, i.e. $\F_\L^- = (\F_\L^\circ)^-$, by virtue of \Cref{lem:generalizedinv_prop}\ref{version}. 

    The pushforward of $\mu^\lambda$ under $\F_\L\circ \L$ is computed by proceeding with the same computations as before after seeing that
    \begin{align*}
        (\F_\L\circ \L)_\#\mu^\lambda ([0,\alpha)) &= \frac{\mu( \F_\L\circ \L< \alpha \text{ and } \L > \lambda)}{\mu(\L > \lambda)}\\
        &= \frac{\F_\L(\F_\L^-(\alpha)-) - \F_\L(\lambda)}{1- \F_\L(\lambda) }.
    \end{align*}
    The formula for $X$ follows from the following straightforward computation 
    \begin{align*}
        (X\circ \L)_\#\mu([0,\beta]) &= 1- (\F\circ \L)_\#\mu ([0,1-\beta))\\
        &=1 - \begin{cases}
            \alpha_i  &\text{ if } 1-\beta \in (\alpha_i, \alpha_i+\Delta_i]\\
            1-\beta & \text{else}
        \end{cases}\\
        &=\begin{cases}
            \beta_i + \Delta_i  &\text{ if } \beta \in (\beta_i, \beta_i+\Delta_i]\\
            \beta & \text{else}
        \end{cases}
    \end{align*}
    where we used the relations $1-\alpha_i = \beta_i+\Delta_i$ and $1-(\alpha_i+\Delta_i) = \beta_i$.
\end{proof}
We now present \Cref{alg:grad} which is the modification of the original Nested Sampling algorithm as proposed in \cite{fowlie2020nested}, but where we split the ``sample generation'' and the ``weight contraction/integration'' computations into separate loops. This does not increase the complexity of the algorithm (also, the weight contraction is purely deterministic and could even be precomputed beforehand), but allows for more modularity which we will use for rare event estimation where we swap out the integrand $\L$.

\newlength{\commentWidth}
\setlength{\commentWidth}{7cm}
\begin{algorithm}[H] 
\begin{algorithmic}
\caption{Modified Nested Sampling according to \cite{fowlie2020nested}} \label{alg:grad}
\STATE{\textbf{Data:} {integrand $\L$, probability measure $\mu$, number of live particles $J$, number of iterations $N$}}
\STATE{\textbf{Result:} {estimate $Z \approx \int \L\diff\mu$, ordered list of dead samples $R$.}}
    \STATE{Generate $J$ samples from $\mu$, call them live particles}
    \STATE{Call $P$ set of live particles}
    \STATE{$R \leftarrow \emptyset$ (Initialize set of dead samples)}
    \STATE{$i \leftarrow 0$}  
    \WHILE{$i < N$ (Sample generation loop)}
        \STATE{$\L^\star \leftarrow \min_j\{\L(p^j), \text{ live particles } p^j\}$}
        \FORALL{live particle $p^j$ such that $\L(p^j) = \L^\star$}{
            \STATE{$i \leftarrow i + 1$}
            \STATE{$R[i] \leftarrow p^j$ (Next element in list of dead samples)}
            \STATE{$P \leftarrow P\setminus p^j$ (Remove particle from live set)}   
        }
        \ENDFOR
        \STATE{Generate enough samples from $\mu$, conditioned to $\L > \L^\star$, in order to fill $P$ to size $J$ again.}
    \ENDWHILE
    \STATE{$Z\leftarrow 0$ (Initialize quadrature)}
    \STATE{$i \leftarrow 0$}   
    \WHILE{$i < N$ (Quadrature loop)}
        \STATE{$i \leftarrow i+1$}
        \STATE{$x_i\leftarrow R[i]$ (Get $i$th dead sample)}
        \STATE{$\xi_i \leftarrow \left(1 - \exp(-1/J)\right) \cdot \exp(-(i-1)/J)$ (deterministic estimate of contraction)}
        \STATE{$Z\leftarrow Z + \xi_i\cdot \L(x_i)$ (Increment quadrature)}
    \ENDWHILE
    \end{algorithmic}
\end{algorithm}
\begin{remark}
    This algorithm has a remarkable property: The selection process which generates the list of dead samples $R$ does not depend on the actual magnitude of the integrand function evaluations, but just on the structure of the level sets of $\L$. Clearly we could replace $\L$ by $\L + c$, or $\L\cdot a$, and keep the same set of dead samples which will be valid dead samples for this modified function, too. Even more, we can replace $\L$ by a completely different function under the condition that their level sets are isomorphic. We make this more rigorous below.
\end{remark}

  We note that \cite{fowlie2020nested} proposed to approximate the unknown quantities $X_i = \mu(\L > \L^\star)$, where $\L^\star$ is the minimum $\L$ value in the live point ensemble at iteration $i$. This is an unknown quantity, but it is being approximated by a deterministic estimator, which is obtained by contracting $X_i$ by $e^{-1/J}$ each time we discard an element of the set of live points. 
This gives rise to the following unproven hypothesis, which we supplement with a sketch of validity in a specific example. 

\begin{hypothesis} \label{hypo}
    We consider the setting of \Cref{lem:formofXL}, i.e. we have an integrand $\L$ such that 
    \begin{align*}
        (X\circ \L)_\#\mu &= \mathrm{Unif}\left(\bigcup_i [\beta_i,\beta_i+\Delta_i)\right)^c + \sum_i \Delta_i \cdot \delta_{\beta_i}.
    \end{align*}
    is a measure on $[0,1]$.

    We further set $\L^\star = \min_j \L(x^j)$, i.e., have that  $X(\L^\star) = \max_{j=1}^J X(\L(x^j))$. 
     Then we can approximate $X(\L^\star)$ deterministically by $X(\L^\star) \approx e^{-n/J}$, where $n = \operatorname{Card}(\{x^j: \L(x^j) = \L^\star\})$.

\end{hypothesis}
\begin{proof}[Sketch of the validity of \Cref{hypo}]
\begin{figure}
    \centering
    \includegraphics[width=\textwidth]{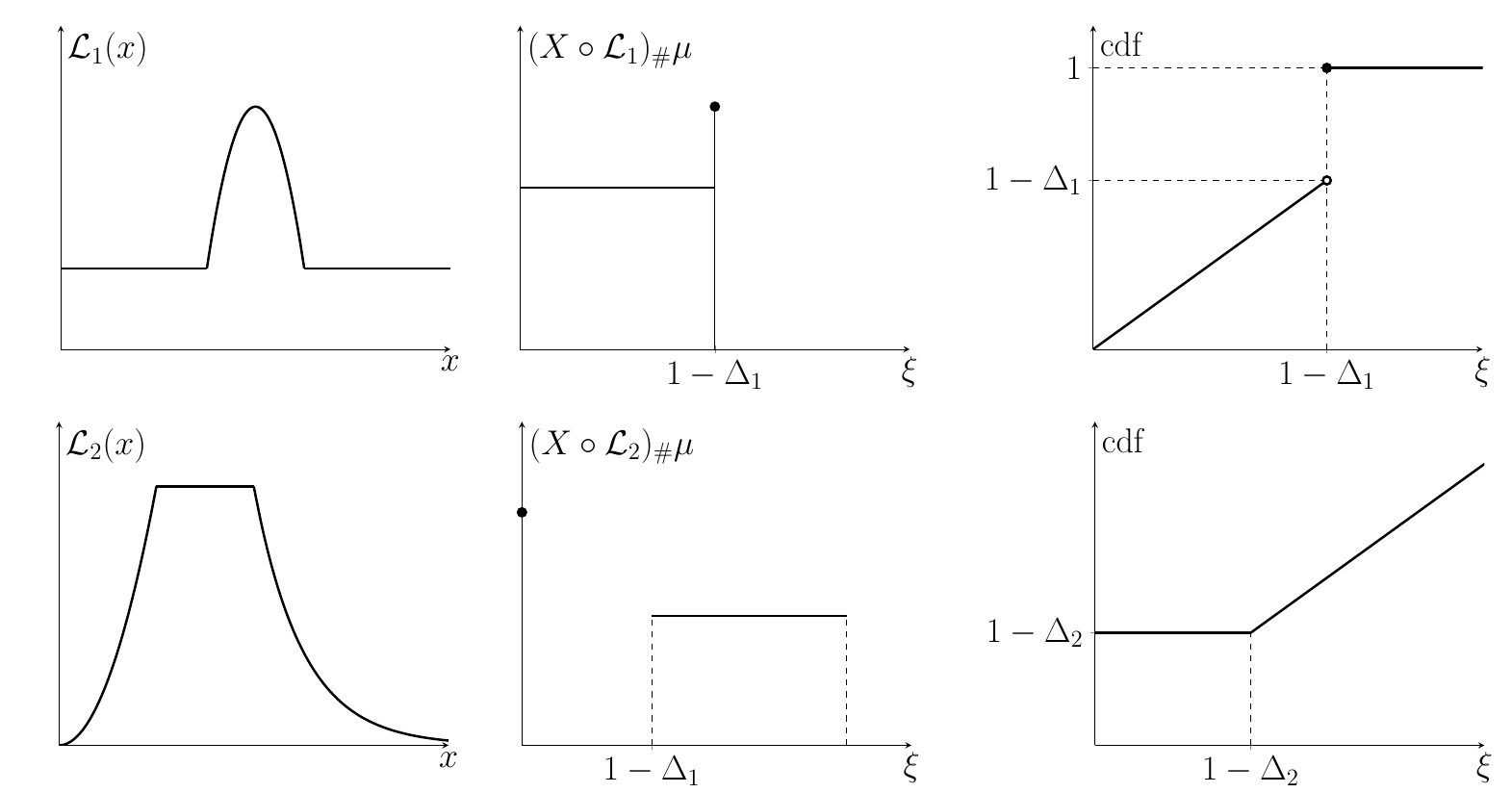}
    \caption{$\L_1$ with a plateau at its minimum, and $\L_2$ with a plateau at its maximum.}
    \label{fig:example_like}
\end{figure}

We consider the two examples in \Cref{fig:example_like}. We denote them by $\L_1$ and $\L_2$. We assume that the measure $\mu$ is absolutely continuous so that indeed $\mu(\L_1=\inf\L_1) = \Delta_1$, $\mu(\L_1 > \inf \L_1) = \beta_1 = 1-\Delta_1$ and $\mu(\L_2 = \sup \L_2) = \Delta_2$, $\mu(\L_2 > \sup\L_2) = \beta_2 = 0$. Hence,
\begin{align*}
    (X\circ \L_1)_\#\mu &= \mathrm{Unif}[0, 1-\Delta_1) + \Delta_1\cdot \delta_{\beta_1}\\
    (X\circ \L_2)_\#\mu &= \Delta_2\cdot \delta_{0}+\mathrm{Unif}[\Delta_2, 1]  
\end{align*}
and the cdfs can be written down as follows:
\begin{align*}
    \mu\left(X\circ \L_1\leq \alpha\right) &= \begin{cases}
        \alpha:&\text{ if } \alpha < \beta_1\\
        1&\text{ if } \alpha \geq \beta_1
    \end{cases}\\    
    \mu\left(X\circ \L_2\leq \alpha\right) &= \begin{cases}
        \Delta_2&\text{ if } \alpha \leq \Delta_2\\
        \alpha:&\text{ if } \alpha > \Delta_2
    \end{cases}
\end{align*}
The cdf of the maximum of $J$ i.i.d. samples is just the $J$-th power of their individual cdf, hence
\begin{align*}
    \mu\left(\max_J X(\L_1(x^k)) \leq \alpha\right)&= \begin{cases}
        \alpha^J:&\text{ if } \alpha < \beta_1\\
        1&\text{ if } \alpha \geq \beta_1
    \end{cases}\\    
    \mu\left(\max_J X(\L_2(x^k)) \leq \alpha\right) &= \begin{cases}
        \Delta_2^J&\text{ if } \alpha \leq \Delta_2\\
        \alpha^J:&\text{ if } \alpha > \Delta_2
    \end{cases}
\end{align*}

We start with the first example, $\L_1$, and we more specifically demand that $J\gg 1$, $\Delta \gg 0$ (or more generally $(1-\Delta)^J\ll 1$). This means that with probability $1-(1-\Delta)^J$, the maximum will be attained at $1-\Delta$.  Nested Sampling employs a contraction of $e^{-1/J}$ for every particle $X(\L_1(x^i)) = \max_j X(\L_1(x^j)) =: \xi$, leading to a total contraction of $e^{-n/J}$, where $N = {\rm card}\{x^j: X(\L_1(x^j)) = \xi\}$. $N$ is binomially distributed: $B\sim {\rm Bin}(J, \Delta)$. This means that the expected number of particles all being the joint maximum is $\E N = J\cdot\Delta$. Ignoring the (very improbable) possibility that the maximum is not attained at $1-\Delta$, the following happens during the foreach loop of considering this plateau: The accessible prior volume is contracted by $e^{-1/J}$ a number of $n = J\cdot \Delta$ on average, i.e. on average, it will have been reduced to $X_n = e^{-\Delta}$, and the evidence will have been increased (telescopic sum!) to $Z = (1 - e^{-\Delta})\L^\star$. This is exactly what we would expect: The plateau has prior contribution $\Delta$, which is roughly equal to $(1 - e^{-\Delta})$ (at least for $\Delta$ not too large), with value $\L^\star$. This correctly (to first order in $\Delta$) handles the plateau, and then Nested Sampling proceeds with the rest of the integrand which does not have any plateaus (and thus this is reduced to the known case).

Now to the second case where $(1-\Delta)^J\gg 0$, i.e., the maximum is with very high probability not attained in the Dirac component. Then calculations are even easier: $N = 1$ and $\xi$ is Beta distributed with mean $\frac{J}{J+1}(1-\Delta)$, which matches the contraction $\exp(-1/J)$ proposed by Nested Sampling to first order (in $J^{-1}$).

Second example: The probability that the maximum is attained in the (left) Dirac is given by $\Delta^J$. We can again distiguish two cases: Case I ($\Delta^J \approx 0$) is again straightforward and we obtain the usual contraction rate. Case II: $\Delta^J \ll 0$ means that there is a nontrivial chance that the maximum (and hence all particles) land in $x=0$. This means that the contraction rate is $\left(\frac{J}{J+1}\right)^J\simeq \exp(-\frac{J}{J+1}) \simeq 1$, which is indeed correct since the contraction is the full interval.
\end{proof}

\section{Application to Rare Event Estimation and beyond}\label{sec:UQ_REE}
In this section we mainly demonstrate how Nested Sampling can be used to efficiently compute rare event probabilities: to this end, we introduce our methodology of surrogate functions that we can also employ to compute moments and to approximate cumulative distribution functions, which we come back to in \Cref{subsec_survhighmom}. The estimation of probabilities of rare events is central in many disciplines, especially in civil engineering.  Nested Sampling in rare event estimation has previously been discussed by \cite{fowlie2022nested,walter2017point}. Especially the fact that Nested Sampling is usually implemented to work with logarithmic quantities only allows for a very stable computation. More traditional methods for rare event estimation include FORM \cite{Rackwitz1978StructuralRU}, subset simulation \cite{AU2001}, sequential importance sampling \cite{PAPAIOANNOU2016}, and the cross-entropy method \cite{RUBINSTEIN1997}. We now introduce our Nested-Sampling-based rare event estimator.

We consider now the problem of computing 
\begin{equation}
    \int_\Omega \chi_A \diff\mu \label{eq:rareEventIntegration}
\end{equation}
where $A$ is a rare event, i.e. we assume that $\mu(A)$ is too small for standard Monte-Carlo estimation. In the previous section we have seen that it is possible to compute this integral via Nested Sampling, even though the integrand $\chi_A$ consists solely of plateaus. However, this would be inefficient: the worker is stuck in the first while-loop of \Cref{alg:grad} until sampling $J$ particles that are in the rare event. To solve this problem, we now propose a variant of Nested Sampling to compute this object in practice. The key idea will be to swap out a regularization of the characteristic function by a level-set surrogate, which is a different function having matching  super-level sets.  The level-set surrogate can be chosen such that it allows for more efficient re-sampling than $\chi_A$ during Nested Sampling. The samples and contractions obtained via the level-set surrogate can be employed to calculate $\int_\Omega \chi_A \diff\mu$ afterwards.


\subsection{A unifying computational trick: Surrogate Integrands}

\begin{definition}[Level-set Surrogate]
    Let $f:\Omega\to\R$ be a continuous mapping. We call $g$ a level-set surrogate for $f$, if for all $\lambda \in\R$ there exists a $\kappa\in\R$ such that $\{x \in \Omega: f(x)>\lambda\} = \{x \in \Omega: g(x) > \kappa\}$.
\end{definition}

\begin{lemma}\label{lem:levelSetSurrogate}
    Let $\{x_i\}_{i=1}^N$, $\{\xi_i\}_{i=1}^N$ be the set of dead samples and estimated contractions obtained from one run of Nested Sampling for the computation of $\int f\diff\mu$. In particular, 
    \[ \int f\diff\mu \approx \sum_{i=1}^N f(x_i) \xi_i.\]
    If $g$ is a level-set surrogate for $f$, then we can use the same set of dead samples to approximate $\int g\diff\mu$:
    \[ \int g\diff\mu \approx \sum_{i=1}^N g(x_i) \xi_i\]
\end{lemma}\begin{proof}
    From \Cref{sec:nestedsampling}, we already know that the estimation of the contraction $\{\xi_i\}_{i=1}^N$ is independent of the integrand. We further know that $g$ is a level-set surrogate of $f$ and, thus, it holds $\argmin_k f(s^k) = \argmin_k g(s^k)$ for any i.i.d sample set $\{s^k\}_{k=1}^J\sim \mu^{\lambda_i}$. Therefore, the lemma follows directly as long as it is ensured that the process of dead sample selection in  Nested Sampling is unchanged by replacing $f$ with $g$.
\end{proof}

We can exploit this observation in at least two ways:
\begin{itemize}
    \item Instead of computing an integral of form $\int \chi_A \diff\mu$, we can construct an integrand $g$ with the property that there exist a value $\kappa$ such that $A =\{g > \kappa\}$. Then $g$ is a level-set surrogate for $\chi_A$ and we can construct samples $\{x_i\}$ and contractions $\{\xi_i\}$ using Nested Sampling performed with $g$. For example, if the integration domain is one-dimensional and $A = (3,\infty)$, then $g(x) = x$ is a valid surrogate function (and Nested Sampling tends to work much better with this integrand since it avoids computational problems associated with plateaus).
    \item If there are multiple integrals of interest with the property that all integrands are level-set surrogates for each other, then we can only need to compute samples and contractions once, and plug them into each integrand. This will be used for the simultaneous computation of moments as well as for the approximation of distribution functions. A further example would be the simultaneous computation of the probabilities of a sequence of nested rare events.
\end{itemize}

Therefore, \Cref{lem:levelSetSurrogate} tells us how to handle ``demanding" integrals more efficiently: we can approximate the calculation of $\int_\Omega \chi_A \diff\mu$ by the calculation of the integral of an appropriate regularized level-set surrogate.\footnote{Actually, it is already sufficient to perform Nested Sampling employing the level-set surrogate until iteration $K$ which is the first iteration in which the removed particle is a rare event ($x_K \in A$). Then, the integral can be approximated by $1 - \sum \xi_K$.} This has mainly the benefit that the computational effort of re-sampling from $\mu^{\lambda_i}$ is reduced due to the fact that an appropriate level-set surrogate allows for a more directed search of new samples within the ``exploration'' method (needed for the generation of new samples satisfying $\L > \lambda^\star$), e.g., gradient based methods.

\textbf{Example:} We consider the case of $\mu = \mathcal N(0,1)$ and $\L = \chi_{[a,\infty)}$. So $\int \L d\mu = \mathcal N(0,1)[a,\infty)$ for $a\gg 1$ is the probability of the rare event of obtaining a large ($\geq a$) positive number when sampling from a standard normal distribution. Then a suitable surrogate integrand is $g(x) = x$ as $\L$ can be approximated arbitrarily well (with $h\to 0$) by a sigmoid function of form $f(x) = \tfrac{1}{2}\tanh((x-a)/h)+\tfrac{1}{2}$, and level sets of $f$ are isomorphic to level sets of $g$.

\subsection{Survival functions and higher-order moments} \label{subsec_survhighmom}

In addition to the rare event probability, we can also infer moments and the cumulative distribution function from the dead samples with (almost) no additional computational cost.
\paragraph{Calculation of moments}
The $k$-th moment is given by $m_k =\int_\Omega x^k \cdot \L(x) \diff \mu(x)$. Here, we can use the particles $\{x_i\}_{i=1}^N$ which were removed from the set of particles in iteration $i$ to approximate this integral:
\begin{align*}
    m_k= \int_\Omega x^k \cdot \L(x) \diff \mu(x)
    \approx \frac{1}{Z} \left({x}_i\right)^k \cdot \L({x}_i) \cdot \left(X(\lambda_{i-1}) - X(\lambda_{i}) \right).
\end{align*}
In fact, this reduces to a weighted sum of the removed particles with normalized weights $\frac{1}{Z} w_i$ with $w_i = \xi_i \cdot \L({x}_i)$, see also \cite{skilling2006nested}. The quantities ${x}_i, \L({x}_i)$ and $w_i$ are already computed while iterating, see \Cref{alg:grad}. Therefore, the calculation of moment $m_k$ requires almost no additional computational effort. We know that the higher the moment, the higher the contribution of samples which are ``far away in the region of the rare event''. This is relativized by the weights $w_i$. Nevertheless, the precision of the moments decreases with increasing $k$.

\paragraph{Approximation of a cumulative distribution function} The cdf can be approximated by the empirical cdf derived from the $N$ weighted particles $\tilde{x}_i$:
\begin{align*}
    F(t) \approx \hat{F}_N(t) = \sum_{i=1}^N \left(X(\lambda_{i-1}) - X(\lambda_{i})\right) \cdot \L(\tilde{x}_i) \cdot \chi_{\tilde{x}_i \le t}(t),
\end{align*}
where again $w_i = (X(\lambda) - X(\lambda)) \cdot \L(\tilde{x}_i)$ is the weight of sample $\tilde{x}_i$. As for the moments, the quantities $\tilde{x}_i, \L(\tilde{x}_i)$ and $w_i$ are already computed while iterating. Therefore, almost no additional effort is spent on the approximation of the cdf after performing Nested Sampling.

\section{Numerical Experiments}\label{sec:simulations}

\subsection{Elementary one-dimensional densities}
We consider first the elementary example of computing the rare event probability $P_a = \mu_i([a,\infty))$ in two settings: $\mu_1$ being a standard Gaussian probability measure, and $\mu_2$ being a Cauchy probability measure. The results of our implementation with Nested Sampling can then be directly compared to the ground truth solution given by $P_a = 1-\Phi(a)$, where $\Phi$ is the Gaussian cumulative distribution function, and similarly for the Cauchy setting. The integrand is thus given by $\L(x) = \chi_{[a,\infty)}(x)$. As a surrogate function, we just use the linear function ${\L_\text{surrogate}}(x) = x$, as motivated earlier.

\begin{figure}
    \centering
    \includegraphics[width=0.5\textwidth]{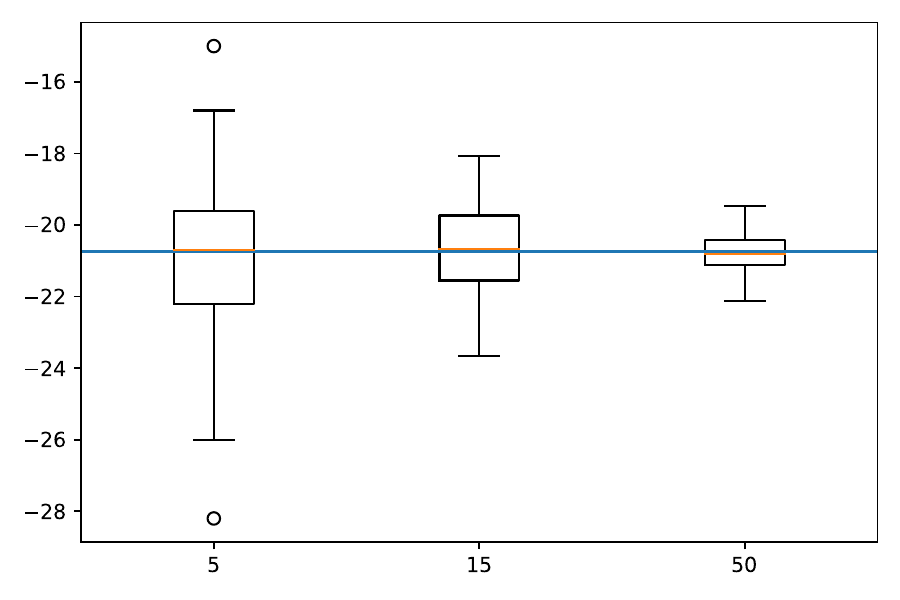}%
    \includegraphics[width=0.5\textwidth]{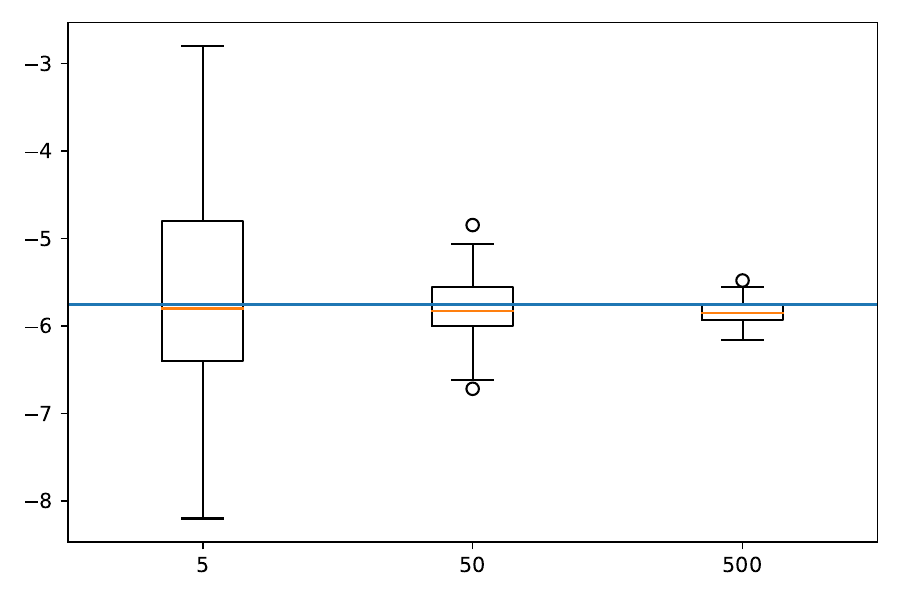}\\
  \includegraphics[width=\textwidth]{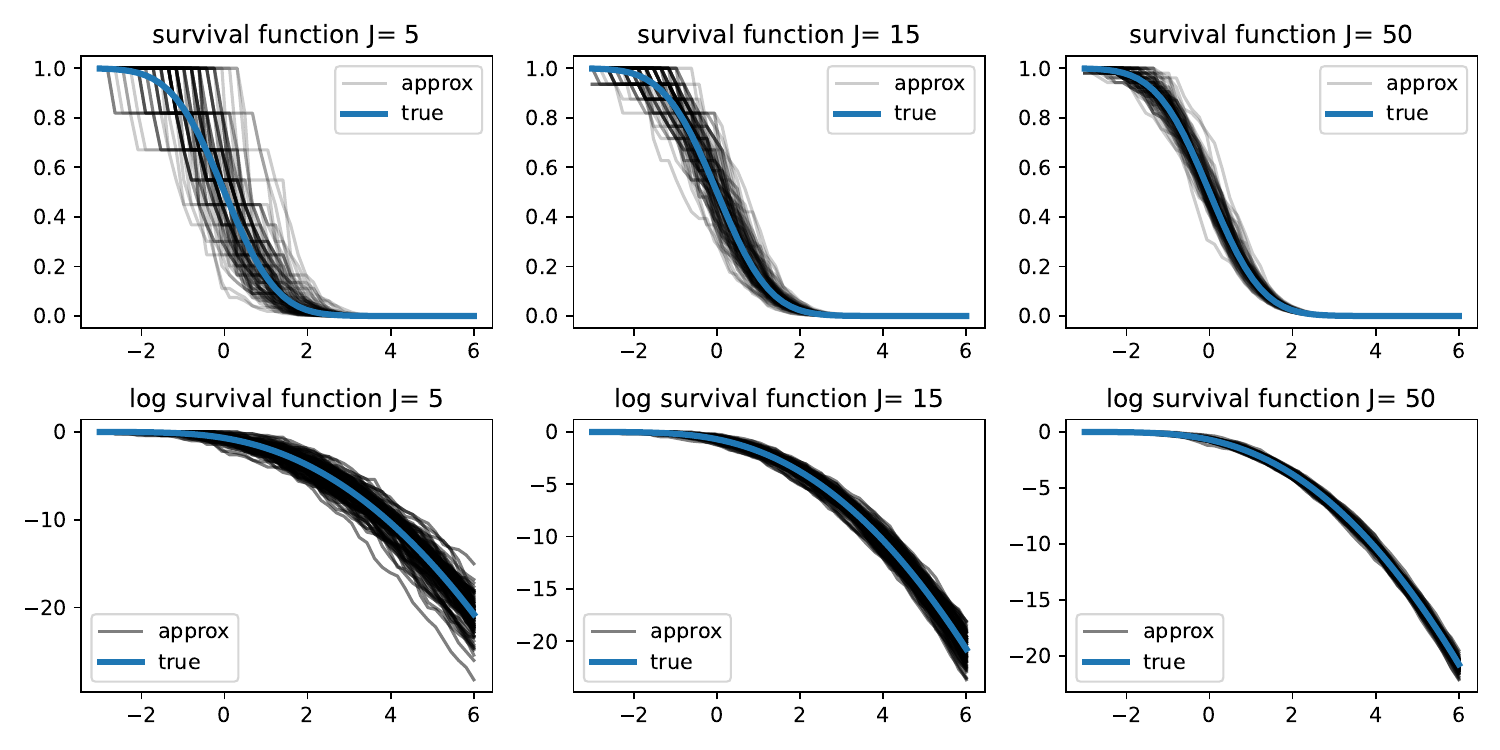}
  \includegraphics[width=\textwidth]{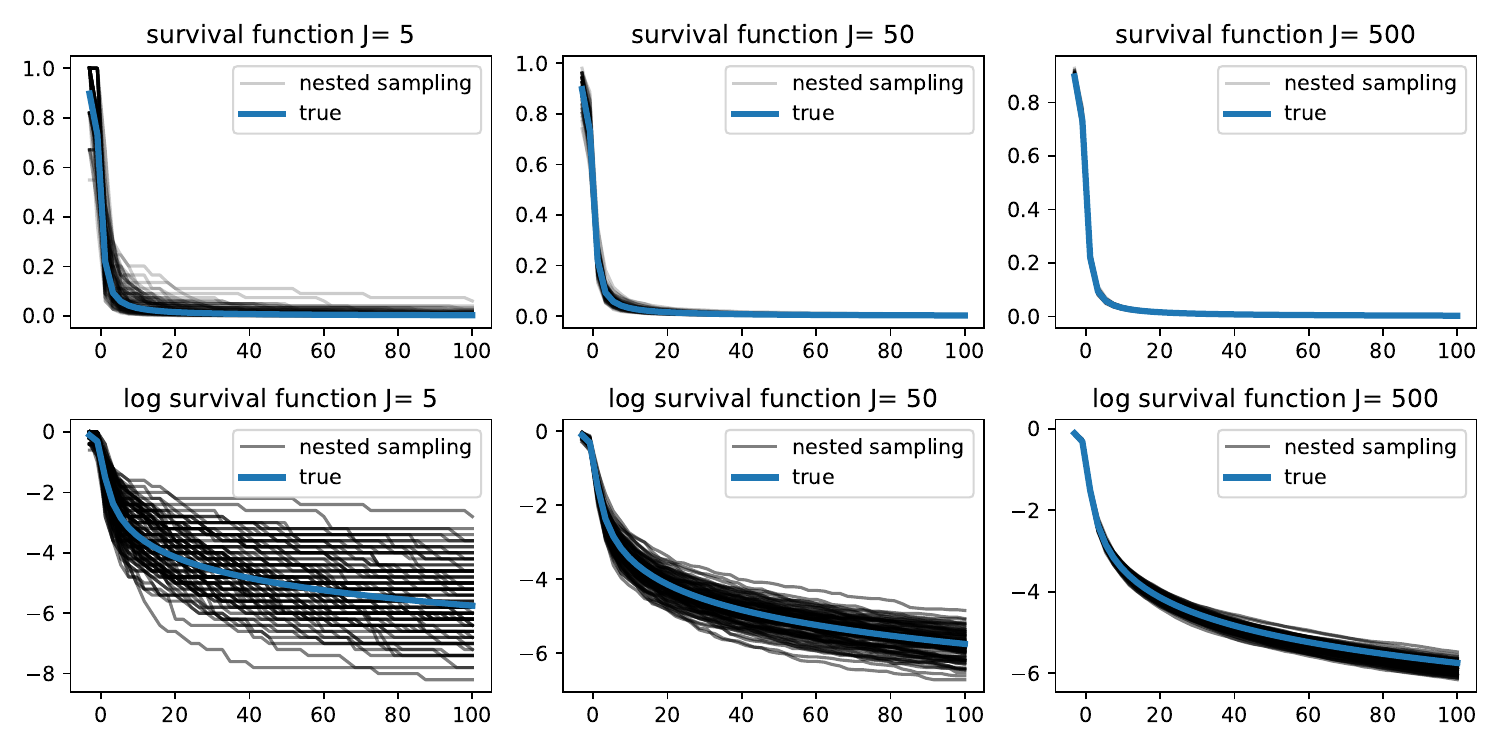}
    \caption{Top row: Estimation of rare event log-probability $\log\mu_i([a_i,\infty))$ for different sizes $J$ of active set. Left: $\mu_1 = \mathcal N(0,1)$, $a = 6$, with $J\in \{5,15,10\}$. Right: $\mu_2 = \mathrm{Cauchy}$, $a=100$, with $J\in \{5,50,500\}$. True value is shown as a blue line. Second and third row: True and estimated (via Nested Sampling) (log-)survival function of $\mathcal N(0,1)$. Bottom two rows: True and estimated (via Nested Sampling) (log-)survival function of a Cauchy probability measure.}
    \label{fig:surv_Gaussian}
\end{figure}

\paragraph{1d standard Gaussian: $P_6$ and moments}
The log probability of the rare event $X>6$ with standard Gaussian random variable $X$ is $\log P_6 \approx -20.74$. We approximate it via Nested Sampling with $J \in \{5, 15, 50\}$ particles. The estimated logprobabilites of 200 independent runs are summarized in \Cref{fig:surv_Gaussian}. Additionally, we give the approximation of the (log)-survival function in \Cref{fig:surv_Gaussian} which is achieved almost without additional computational effort.

\paragraph{1d Cauchy: $P_{100}$ and cdf}
The probability of the rare event $X>100$ with Cauchy random variable $X$ is $\log P_6 \approx -5.75$. The approximation results obtained from Nested Sampling using $J \in \{5, 50, 500\}$ particles are shown in \Cref{fig:surv_Gaussian} and the approximation of the (log-)survival function is shown in \Cref{fig:surv_Gaussian}.

\subsection{Diffusion in a double-well potential}
We consider a double-well potential $V(x) = -\frac{a}{2}x^2 + \frac{b}{4}x^4$, where we set $a=2$ and $b = 0.5$. Then $V$ has two wells at $\pm \sqrt{a/b} = \pm 2$.
We consider the overdamped Langevin dynamics of a particle (initialized to start in one of the wells) in this potential governed by
\begin{equation*}
    \diff x_t = -\nabla V(x_t) +\sigma \diff W_t,\quad x_0 = 2.
\end{equation*}

\begin{figure}
    \centering
    \includegraphics[width=0.49\textwidth]{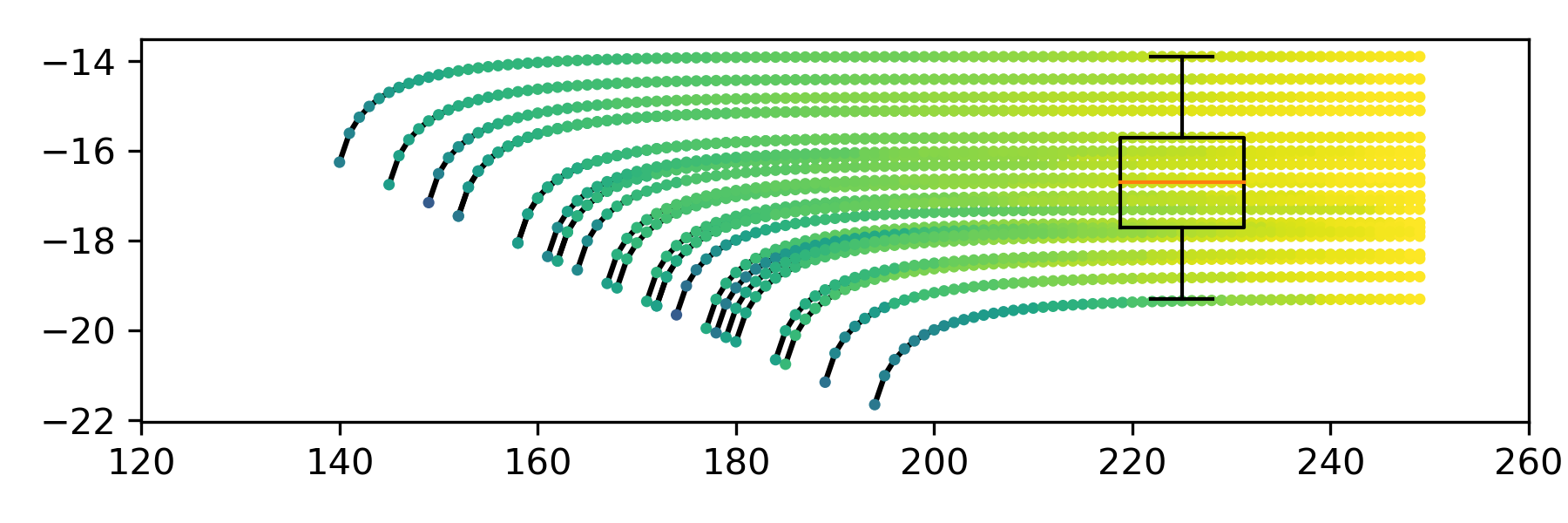} 
    \includegraphics[width=0.49\textwidth]{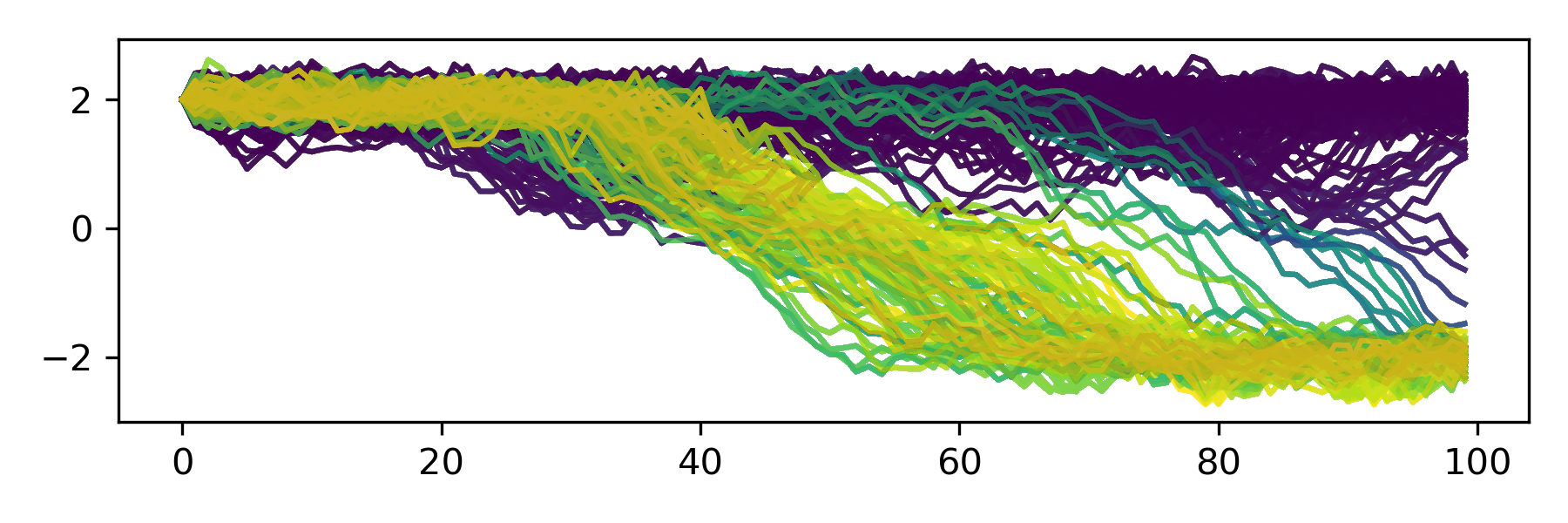}\\
    \includegraphics[width=0.49\textwidth]{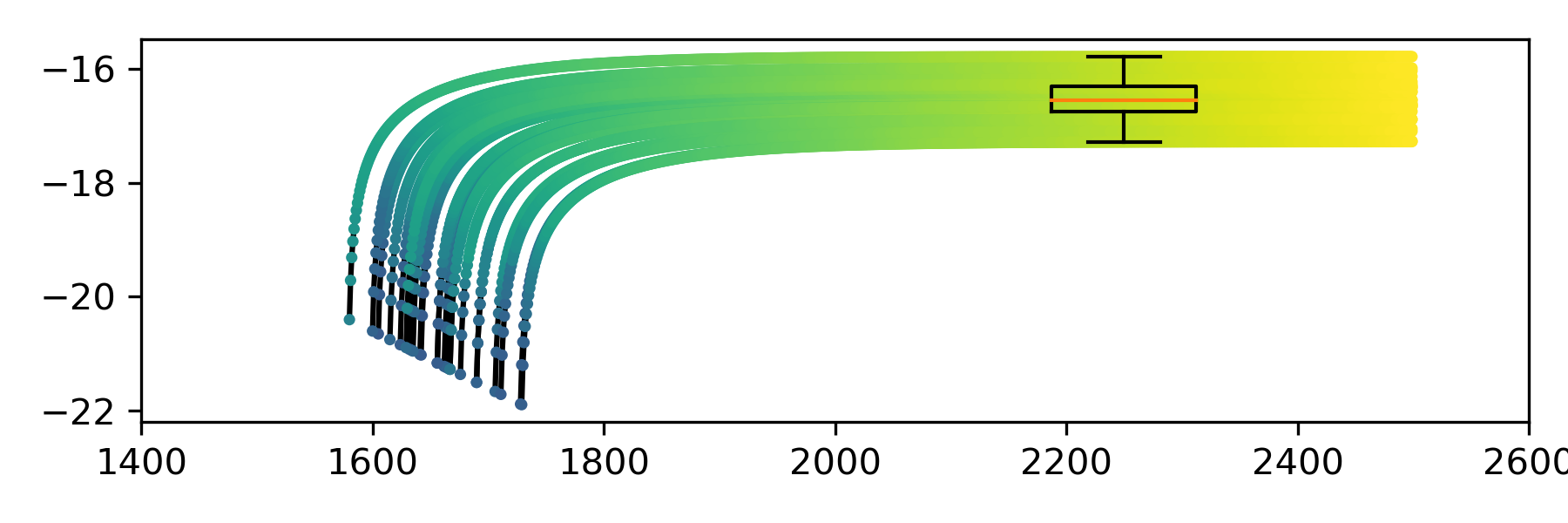}  
    \includegraphics[width=0.49\textwidth]{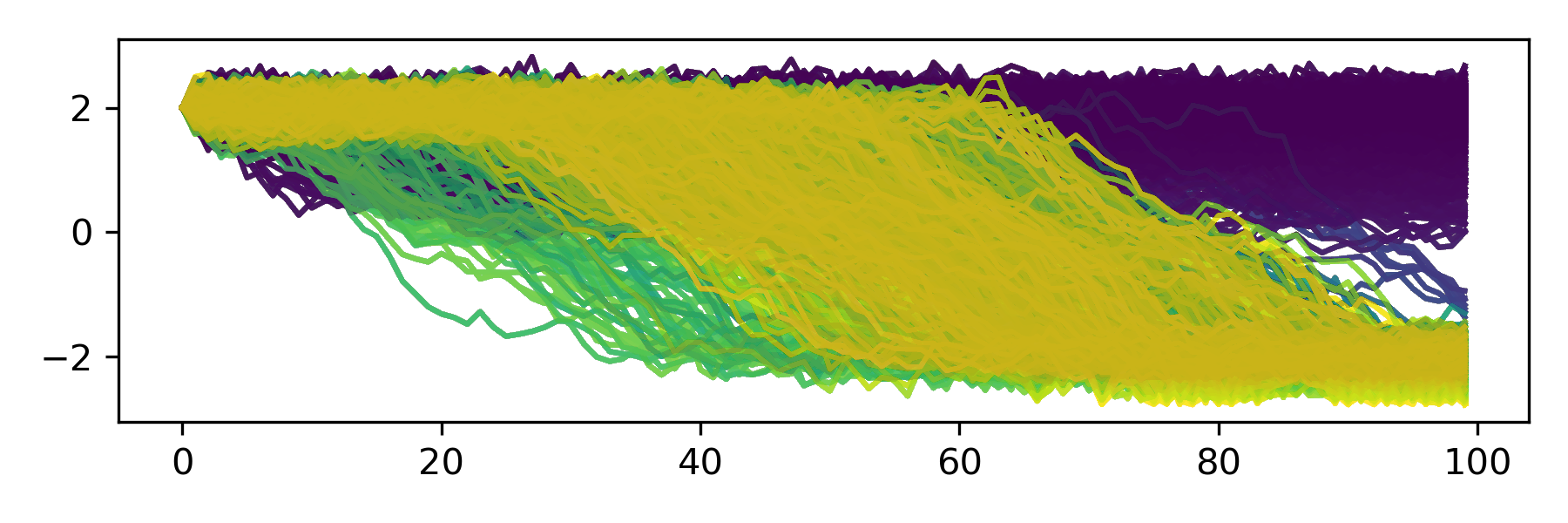}\\%
    \caption{Rare Event estimation for double-well diffusion. Top row: $J=10$. Bottom row: $J=100$. Left column: accumulated value of $\log Z$ during 25 independent runs of Nested Sampling (boxplot shows distribution of final value of computed rare event probability). Right column: Visualization of samples (i.e. paths in double well potential) removed from active set along with iteration of Nested Sampling. Color of path varies with iteration from dark (early) to light (late). }
    \label{fig:doublewell}
\end{figure}

We are interested in the probability of the path $\{x_t\}_{t\in[0,T]}$ leaving its initial well at $x_0=2$ within $t\in[0,10]$. This is a rare event, and due to the attractive nature of the second well we can specify it via $A=\{\min_{t\in[0,T]}x_t \leq -1.5\}$, i.e. $\mu(A) = \int_\Omega \chi_A(\{x_t\}_t) \diff\mu(\{x_t\}_t)$ if $\mu$ is the distribution of the stochastic process $\{x_t\}_t$. As a level-set surrogate function we can choose  $g(\{x_t\}_t) = \min_{t\in [0,T]}x_t$. \Cref{fig:doublewell} shows the evolution of accumulated log-probability of $25$ independent runs of Nested Samplings for ensemble sizes $J=10$ and $J=100$ each, as well as (for one of the $25$ runs) a visualization of all discarded samples (which correspond to specific diffusion paths) created during a specific run of Nested Sampling. Darker paths in the right part of \Cref{fig:doublewell} correspond to typical (prior-like) simulation outcomes, with lighter paths corresponding to (rarer) paths closer to the rare event of switching to the secondary potential.

It can be observed that a larger number of ensemble members $J$ decreases the variance of the Nested Sampling estimator of the log probability, but requires a higher number of iterations until convergence.
\subsection{Loaded beam with random inclusions} We now study a rare event estimation problem that arises in an engineering context. We consider a beam (1D) that consists of a generally homogeneous material that is subject to small inclusions. These inclusions affect the beam's flexibility $F$. 
We now affix the beam horizontally on one side and attach a small point load on the other side. Given a stochastic model for the inclusions, we are interested in the probability of the beam vertically deflecting beyond a certain point.

The beam has length $L$; the flexibility is given by a sum of a constant $c > 0 $ and a Poisson point process on $[0, \infty)$. Then, the flexibility is given as a measure
$$
F(\mathrm{d}t) = c \mathrm{d}t + \sum_{i = 1}^\infty w \delta(\mathrm{d}t-T_i),$$
where $T_i = \sum_{j=1}^i\Delta T_i$,  $\Delta T_i$ are independent exponentially distributed random variables with rate $\lambda>0$ $(i\in \mathbb{N})$, and $w$ is a weight. We choose this model for the flexibility to reflect that the inclusions are spatially extremely small, but have a significant influence. 
The deflection response $d$ is given through an Euler--Bernoulli equation and reads:
$$
d(x) = -P \int_0^x \int_0^s (L-t) F(\mathrm{d}t) \mathrm{d}s  \qquad (x \in [0, L]).
$$
This integral can be solved analytically:
$$
d(x) = -P\left(\frac{cLx^2}{2} - \frac{cx^3}{6} + \sum_{i : T_i \leq x} w(L-T_i)x \right) \qquad (x \in [0, L]).
$$
For our experiment, we consider a beam of length $L=5$, usual flexibility constant $c = 1$, load $P := 0.01$, additional flexibility at inclusions $w := 0.05$, and inclusion rate $\lambda = 1$. We assume that the rare event occurs when the beam deflects beyond the point $-0.55$, which is about 30\% lower than the deflection under no inclusions: $-0.417$.

\begin{figure}
    \centering
    \includegraphics[width=\textwidth]{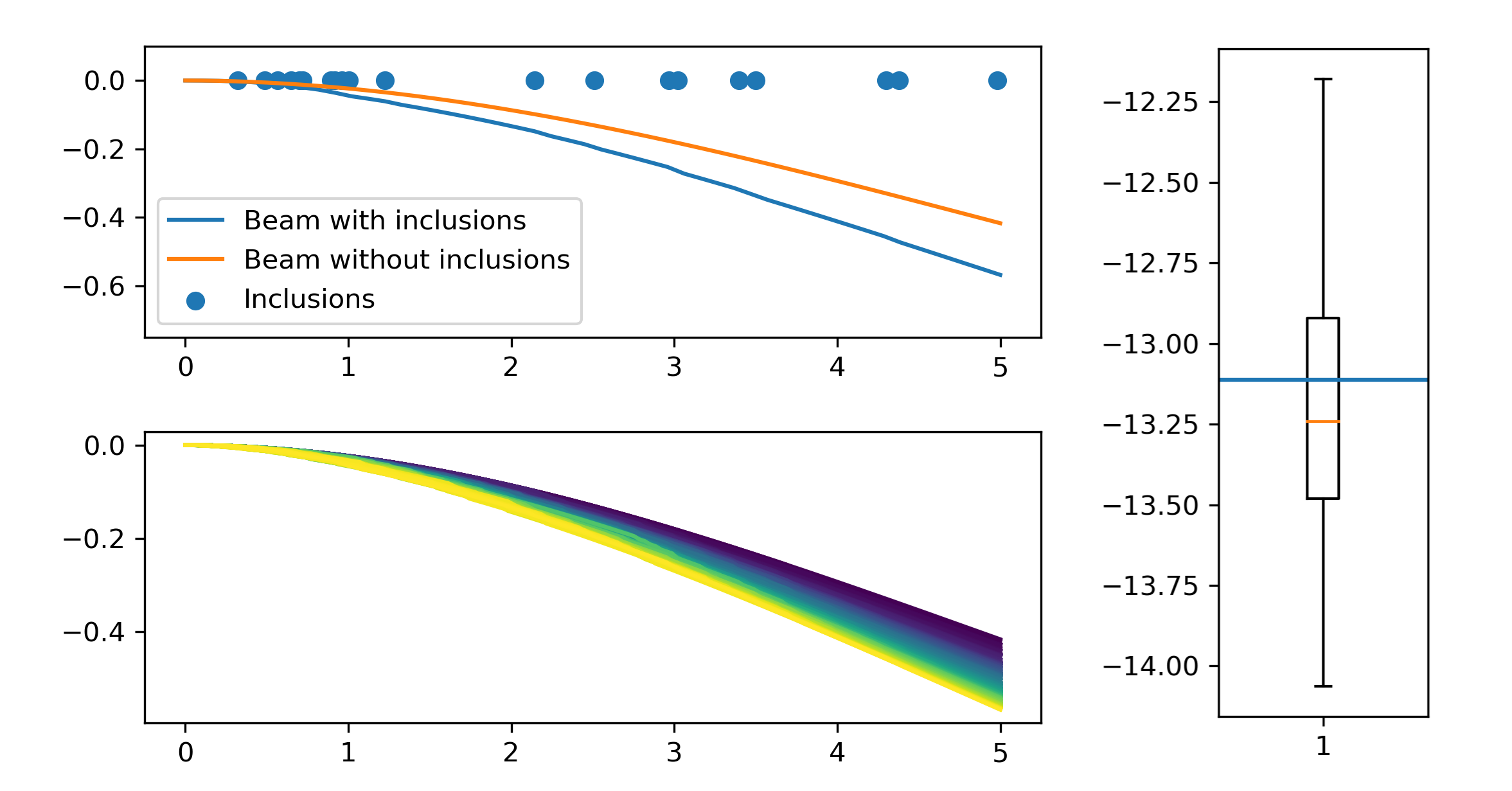}
    \caption{Top: Loaded beam with and without inclusions. Bottom: Samples generated by Nested Sampling during iteration (dark = early, light = late iteration). Right: Logprobability of rare event (deflection exceeding $-0.55$), as computed by 20 independent runs of Nested Sampling (Boxplot), in comparison to brute force Monte Carlo simulation with $10^9$ runs (bootstrapped - horizontal line).}
    \label{fig:beam}
\end{figure}

To discretise the Poisson point process given above, we assume a maximum number of inclusions of $20$; given inclusion rate and length of the rod, the probability to have more than 20 inclusions is very small; we have $\mathrm{Poisson}(5)([21,\infty))  = 8.11\cdot 10^{-8}$. This is two orders of magnitude smaller than the probability of the rare event, which we estimated with $10^8$ Monte Carlo samples and give it in the rightmost panel in ~\Cref{fig:beam}. The top panel shows a beam without inclusions as well one with rare event inclusions leading to displacement beyond the critical threshold. The bottom panel visualizes (similarly to our simulations for the double-well potential) displacements corresponding to samples discarded during the iteration of Nested Sampling. The surrogate function used was the negative displacement of the beam.


\subsection{Random Walk in a labyrinth}
\begin{figure}
    \centering
    \includegraphics[width=\textwidth]{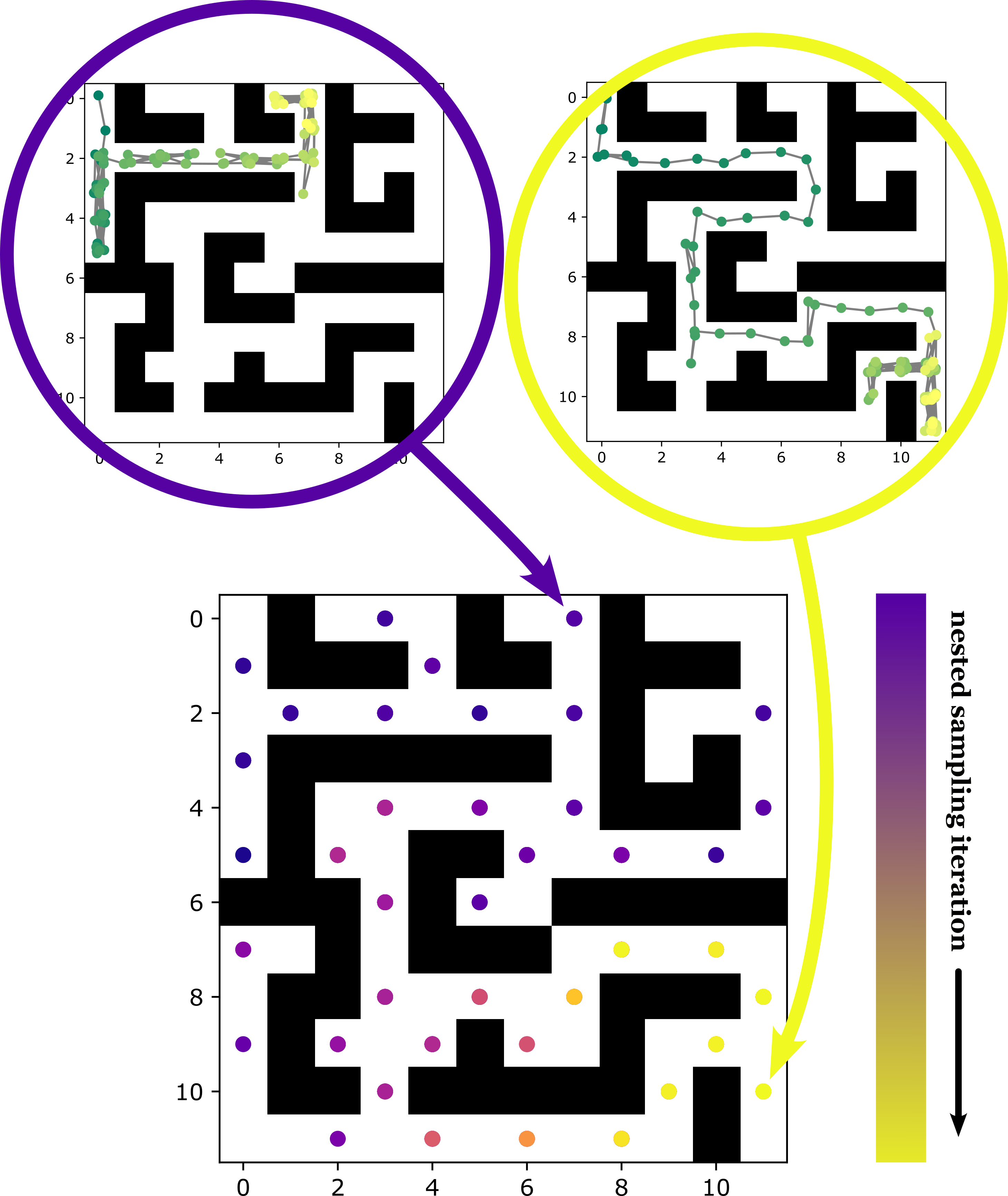}
    \caption{A two-dimensional labyrinth with start $(0,0)$ (upper left corner) and goal $(11,11)$ (lower right corner) and a randomly sampled random walk of length $100$. Main plot below: Each point marks the final position of the random walks sampled from Nested Sampling, with light colors being generated later in Nested Sampling's iteration. Top row: Two sample random walks, early (left) and last (right) in Nested Sampling's iteration.}
    \label{fig:labyrinth}
\end{figure}

We consider a discrete state space $\mathcal S := \{0,\ldots,N-1\}\times\{0,\ldots,N-1\}$ in the form of a two-dimensional grid with ``prohibited states''(walls), forming a labyrinth as in \Cref{fig:labyrinth}. On this labyrinth we set a random walk of length $K$, starting at $(0,0)$. In each step, the random walk considers all four possible directions $\{\text{North},\text{South},\text{East},\text{West}\}$ and chooses a random direction under the condition that it does not end up on a prohibited state or outside of the state space. We want to calculate the probability that this random walk of length $K$ finds the exit, located at $(N-1,N-1)$; i.e. it visits this cell at least once within $K$ steps. Simulations below are for $N=12$, and $K=100$. Writing $w = (w_0,\ldots,w_{K-1})$ with $w_i\in \mathcal S$ for the random walk, and $d: \mathcal S\to \N_0$ for the step-wise distance function of a given cell $s\in \mathcal S$ to the exit, the surrogate function used was $g(w) = -\min_{i\in \{0,\ldots,K-1\}}\{d(w_i)\}$. The method used for finding random walks under the condition of ``distance not exceeding a given threshold'' was rejection sampling (on the space of random walks).

\begin{figure}
    \centering
    \includegraphics[width=0.75\textwidth]{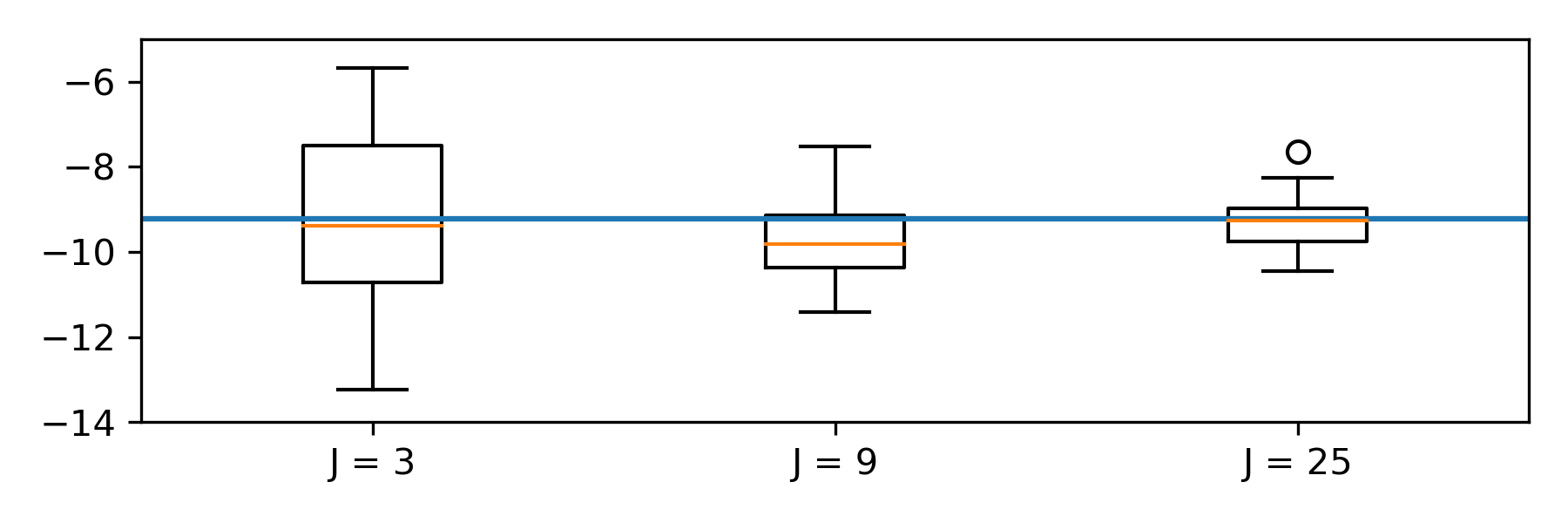}
    \caption{Estimates of logprobability of ``accidental labyrinth solve'' via Nested Sampling, for $J=3$, $J=9$, and $J=25$, with 25 separate runs each. Blue line marks a brute-force estimate of the logprobability obtained by brute force Monte Carlo sampling one million random walks.}
    \label{fig:histogram_labyrinth}
\end{figure}

\section{Conclusions and outlook}
We have studied the generalisation of Nested Sampling towards general quadrature problems, including the plateau case, and have provided both mathematical verification of correctness and computational improvement via the surrogate function methodology. A quadrature problem that we have focused on particularly is that of estimating rare event probabilities. Here, we have shown efficiency and effectiveness of our estimator in a multitude of very heterogeneous numerical experiments.

A rigorous proof of \Cref{hypo}, as well as an accurate estimation of the quadrature error produced by the Nested Sampling procedure are obvious directions for future research.

\appendix

\section{Facts about generalized inverses}
The following lemma can be found, e.g., in \cite[proposition 2.(1)]{embrechts2013note}, but is recorded here for completeness of exposition.
\begin{lemma}\label{lem:uniformmeasure}
    If $\L: \Omega\to \R$ is a measureable map without plateaus of nonvanishing measure $\mu$, and $X$ is its survival function $X(\lambda) = \mu(\L > \lambda)$, then the push-forward of $\mu$ along $X\circ \L$ (or alternatively, the law of $X\circ \L$, interpreted as a random variable) is the uniform measure on $[0,1]$
\end{lemma}
\begin{proof}
    This is easiest seen by looking at the cumulative distribution function of $\F_\L\circ \L$, where $\F_\L = 1-X$ is the cdf of $\L$. We note that $X: \R\to (0,1)$ is invertible since $\L$ has no plateaus
    \begin{align*}
        \mu(\F_\L\circ \L \leq \alpha) &= \mu(\L\leq \F_\L^{-1}(\alpha)) = \F_\L(\F_\L^{-1}(\alpha))= \alpha
    \end{align*}
    This shows that the cdf of $\F_\L\circ \L$ is indeed the cdf of a uniform distribution on $(0,1)$. Since $X = 1 - \F_\L$, the same holds for $X$.
\end{proof}
The following statements are from \cite{wacker2023please}.
\begin{definition}[Generalized inverse]
Let $T:\R\to \R$ be a non-decreasing\footnote{i.e. $x<y$ implies $T(x)\leq T(y)$} function where we set $T(-\infty) = \lim_{x\to-\infty}T(x)$ and $T(\infty)=\lim_{x\to\infty}T(x)$. Then the left-continuous generalized inverse $T^-:\R\to\bar \R$ of $T$ is defined by
\begin{align}
    T^-(y) &= \inf\{x\in \R: T(x) \geq y\}
\end{align}
with the convention that $\inf \emptyset = \infty$.
\end{definition}

We follow up with a list of elementary properties of $T^+$. This is an adaptation of \cite[Proposition 1]{embrechts2013note} to our version of generalized inverse and similar to \cite[Proposition 4.2]{de2015study} (but with some errors fixed).
\begin{lemma}\label{lem:generalizedinv_prop} Let $T:\R\to\R$ be a nondecreasing map.
        \begin{myenum}
            \item \label{equiv_rightcont} $ y \leq T(x)$ if and only if $T^-(y)\leq x$.
        \item \label{version} We define the left-continuous and right-continuous versions $T_l(x):=T(x-)$ and $T_r(x):=T(x+)$ of $T$. Then $T_l^+ = T_r^+$ as well as $T_l^-=T_r^-$.
        \end{myenum}
\end{lemma}

\begin{lemma}\label{lem:TTpmTpmT}
    Let $T$ be nondecreasing and \textbf{continuous from the left}. We denote by $X = \{x_i\}$ the (ordered) list of all discontinuities of $T$, i.e. $y_i^+ := T(x_i+)>T(x_i) =: y_i^-$ and $T(x+) = T(x)$ for $x\not\in X$. We denote by $Y = \{y_i\}$ the (ordered) list of plateaus of $T$, i.e. for each $y_i$ there exists a proper (maximal in the set of half-open intervals) interval $I_i = (x_i^-, x_i^+]$ such that $T(x)\equiv y_i$ for all $x\in I_i$. Then 
    \begin{align*}
    T(T^-(y)) &= \begin{cases}y_i^-,&\text{ for } y\in(y_i^-,y_i^+]\\
    y,& \text{ else }
    \end{cases}\\
    T^-(T(x)) &= \begin{cases}x_i^-,&\text{ for } x\in(x_i^-,x_i^+]\\
    x,& \text{ else }
    \end{cases}
    \end{align*}
\end{lemma}

\section*{Acknowledgments}
 D.S and P.W. want to thank Maria Neuss-Radu for fruitful discussion.

\bibliographystyle{siamplain}
\bibliography{references}
\end{document}

%% file: ex_shared.tex
\usepackage{lipsum}
\usepackage{amsfonts}
\usepackage{graphicx}
\usepackage{epstopdf}
\usepackage{algorithmic}
\ifpdf
  \DeclareGraphicsExtensions{.eps,.pdf,.png,.jpg}
\else
  \DeclareGraphicsExtensions{.eps}
\fi

\newsiamremark{remark}{Remark}
\newsiamremark{hypothesis}{Hypothesis}
\crefname{hypothesis}{Hypothesis}{Hypotheses}
\newsiamthm{claim}{Claim}

\headers{Nested Sampling and Rare Event Estimation}{J. Latz, D. Schneider, and P. Wacker}

\title{Nested Sampling for Uncertainty Quantification and Rare Event Estimation\thanks{Submitted to the editors .
\funding{}}}

\author{Jonas Latz\thanks{Department of Mathematics, University of Manchester, UK
  (\email{jonas.latz@manchester.ac.uk}).}
\and Doris Schneider\thanks{Ostbayerische Technische Hochschule Amberg-Weiden, Germany (\email{d.schneider@oth-aw.de}).}
\and Philipp Wacker\thanks{School of Mathematics and Statistics, University of Canterbury, Christchurch, New Zealand (\email{phkwacker@gmail.com}).}}

\usepackage{amsopn}

\usepackage{enumitem,cleveref}
\usepackage{todonotes}

\renewcommand{\L}{\mathcal{L}}
\newcommand*\diff{\mathop{}\!\mathrm{d}}
\newcommand{\R}{\mathbb{R}}
\newcommand{\N}{\mathbb N}
\newcommand{\F}{\mathcal{F}}
\newcommand{\E}{\mathbb{E}}
\newcommand{\argmin}{\operatorname{argmin}}
\newcommand{\argmax}{\operatorname{argmax}}
\newcommand{\eps}{\varepsilon}

\newtheorem{assumption}{Assumption}
\crefname{assumption}{assumption}{assumptions}
\Crefname{assumption}{Assumption}{Assumptions}

\newlist{myenum}{enumerate}{3}
\setlist[myenum,1]{label={\normalfont(\alph*)},
                   ref  ={\normalfont(\alph*)}}
\setlist[myenum,2]{label=(\roman*),
                   ref  =\themyenumi{.(\roman*)}}
\crefname{myenumi}{}{}
\crefname{myenumii}{}{}

